\documentclass[aps,prd,twocolumn,superscriptaddress,nofootinbib,floatfix]{revtex4-2}
\usepackage{amsmath,amssymb,graphicx,bm,xcolor,booktabs}
\usepackage[colorlinks=true,linkcolor=black,citecolor=black,urlcolor=black]{hyperref}

\newcommand{\TabsRange}{0.1 s to 7 minutes}
\newcommand{\TabsOPE}{0.12--36}
\newcommand{\TabsBase}{0.81--37}
\newcommand{\TabsPP}{11--400}
\newcommand{\TabsMinMassBase}{100}
\newcommand{\TabsMinMassPP}{30}
\newcommand{\EosFacBase}{1.6--3.0}
\newcommand{\EosFacOPE}{3.1--3.4}
\newcommand{\EosFacPP}{10.5}
\newcommand{\npColdTen}{2.4--2.7}
\newcommand{\npColdHi}{17--80}
\newcommand{\npColdHiMed}{23--30}
\newcommand{\RcoldRange}{11.9--13.4}
\newcommand{\DcoldGMfrac}{3}
\newcommand{\DcoldGMchange}{7}
\newcommand{\DcoldFSfrac}{21}

\newcommand{\RcoldGM}{13.60}
\newcommand{\RcoldFS}{12.93}
\newcommand{\RcoldDone}{13.59}
\newcommand{\RcoldDtwo}{12.45}
\newcommand{\DcoldFSfac}{1.5--2.7}
\newcommand{\MmaxGMD}{1.89}
\newcommand{\MmaxFSD}{1.73}
\newcommand{\MmaxGMDone}{2.01}

\newcommand{\PnsRten}{12.5--13.4}
\newcommand{\PnsMgten}{1.34--1.37}
\newcommand{\PnsTpeak}{34--40}
\newcommand{\PnsRb}{23}
\newcommand{\PnsRc}{17}
\newcommand{\DoneCenter}{2.3}

\newcommand{\DoneFirst}{1.0}
\newcommand{\DtwoCenter}{8}

\newcommand{\ppScaleGM}{4.70}
\newcommand{\opeScaleGM}{0.145}
\newcommand{\ppScaleFS}{5.15}
\newcommand{\opeScaleFS}{0.147}
\newcommand{\RLrange}{$9\times10^{-9}$--$6\times10^{-5}$}
\newcommand{\RLspread}{2.6--4.7}
\newcommand{\DecSlow}{0.86--0.91}
\newcommand{\MCerr}{39}
\newcommand{\MCerrMed}{5}
\newcommand{\JmRange}{7.7--196}
\newcommand{\JgRange}{$4.6\times10^{-19}$--$2.0\times10^{-16}$}
\newcommand{\Jmargin}{43}
\newcommand{\JmarginFav}{180}
\newcommand{\JmRangeFav}{4.2--215}
\newcommand{\JnoAbsMax}{$9\times10^{6}$--$2\times10^{7}$}
\newcommand{\JabsMax}{43--180}
\newcommand{\JedgeRise}{280--700}
\newcommand{\JceilFall}{1.9--2.6}
\newcommand{\JedgeTen}{$5.9\times10^{-17}$}
\newcommand{\JedgeHundred}{$4.6\times10^{-19}$}
\newcommand{\ConsFac}{1.64}
\newcommand{\JmRangeCons}{8.9--183}
\newcommand{\JmarginCons}{26}
\newcommand{\baseTwoMargin}{72}
\newcommand{\ppmRange}{9.6--141}
\newcommand{\ppMargin}{10}
\newcommand{\opemRange}{8.3--231}
\newcommand{\opeMargin}{48}
\newcommand{\rateSpread}{7.2}
\newcommand{\RsensRlo}{11.5}
\newcommand{\RsensRhi}{13.5}
\newcommand{\RsensHi}{236}
\newcommand{\RsensMid}{107}
\newcommand{\RsensLo}{45}
\newcommand{\RsensRatio}{5}
\newcommand{\RsensRatioSet}{3.7}
\newcommand{\RsensmHi}{3.9--251}
\newcommand{\RsensmLo}{7.6--187}
\newcommand{\DoneRL}{0.86--0.95}
\newcommand{\DoneMargin}{72}
\newcommand{\BOneMargin}{72}
\newcommand{\DtwoRL}{1.3--2.8}
\newcommand{\DtwoMargin}{312}
\newcommand{\FOneMargin}{180}
\newcommand{\DeltaFac}{1.0--1.7}
\newcommand{\UnionG}{$4.6\times10^{-19}$--$5.7\times10^{-15}$}
\newcommand{\UnionM}{6--213}
\newcommand{\BnineSpinFac}{9}
\newcommand{\TauMin}{$10^{8}$}
\newcommand{\AbsDecayOrders}{5}
\newcommand{\UnionGmax}{$6\times10^{-15}$}
\newcommand{\gindTen}{$2.7\times10^{-15}$}
\newcommand{\gindHundred}{$2.7\times10^{-13}$}
\newcommand{\tuneTen}{14--46}
\newcommand{\tuneHundred}{$3\times10^{4}$--$6\times10^{5}$}
\newcommand{\SatDev}{0.2}
\newcommand{\GMsat}{$n_0=0.153$ fm$^{-3}$, $E_0=-16.4$ MeV, $K_0=241$ MeV, $J=32.5$ MeV and $M^*/M=0.78$}
\newcommand{\ThermoDev}{$4\times10^{-5}$}
\newcommand{\DeltaThermo}{$3\times10^{-7}$}
\newcommand{\EtaGapGM}{0.24--3.0}
\newcommand{\EtaGapFS}{0.55--8.6}
\newcommand{\TabsApoFac}{2.8}
\newcommand{\Ncross}{$6\times10^{18}$}
\newcommand{\PorbMs}{1.9}
\newcommand{\EconsRange}{0.5--2.3}
\newcommand{\DthFermi}{186}
\newcommand{\RtabsCorr}{$-0.88$ to $-0.84$}
\newcommand{\RLrateFac}{3.5}
\newcommand{\PiAbsRatio}{2.8--1700}
\newcommand{\PiHoFac}{15--17}
\newcommand{\PiZeroShare}{53--61}
\newcommand{\PiDeltaShare}{75}
\newcommand{\PiProdFac}{2.5--290}
\newcommand{\PiSurvFac}{5.4--400}
\newcommand{\JmRangePi}{7.7--260}
\newcommand{\JmarginPi}{59}
\newcommand{\JmUpperFavPi}{320}
\newcommand{\PiUpperOthers}{218--272}
\newcommand{\NsetsAll}{257}
\newcommand{\NsetsMass}{165}
\newcommand{\NsetsPass}{19}
\newcommand{\PonsMgOurs}{1.961, 2.093 and 2.046}
\newcommand{\PonsMgDev}{0.8--2.4}
\newcommand{\ConsLeptonLoss}{22}
\newcommand{\LaLnuPeak}{91}
\newcommand{\EaEnuMax}{60}
\newcommand{\LaLnuPeakMass}{100}
\newcommand{\EaEnuTen}{2}

\begin{document}

\title{Absorption of bound supernova axion-like particles by the neutron star weakens the heating limits on their photon coupling}

\author{Probit J Kalita}
\author{Subha Mitra}

\author{Bharat Kumar}

\email{kumarbh@nitrkl.ac.in}
\affiliation{Department of Physics and Astronomy, National Institute of Technology, Rourkela 769008, India}
%% co-authors to be added

\date{\today}

\begin{abstract}
Axion-like particles (ALPs) that couple to protons are produced in a core-collapse supernova, and the slow ones remain bound to the new neutron star. Bose, Ganguly, Kumar, and Raj [arXiv:2610.30255] proposed that the photons from the decay of this bound cloud reheat old neutron stars, and they derived strong limits on the ALP-photon coupling. They neglected the absorption of the ALPs by the star, but a bound ALP on an orbit that enters the star and reaches 50 km crosses it about \Ncross\ times over the 333 Myr age of PSR J2144$-$3933. We show that inverse nucleon bremsstrahlung in the cold star absorbs every ALP whose orbit enters it within minutes, long before it decays, so only ALPs born outside the final star on nearly circular orbits survive; they carry \RLrange\ of the heating. We compute the protoneutron-star evolution, the production, the general-relativistic orbits and the absorption ourselves, for twelve protoneutron-star evolutions, three nucleon-bremsstrahlung rates normalized to the same supernova energy loss of light ALPs, and equations of state with and without $\Delta(1232)$ isobars. PSR J2144$-$3933 still excludes a window at $m_a\simeq$ \JmRange\ MeV, with a lower edge in the photon coupling \JedgeRise\ times higher than without absorption. Pion conversion and the $\Delta(1232)$ resonance, which raise the production of surviving ALPs more than their absorption, widen this window to $m_a\simeq$ \JmRangePi\ MeV. The window depends at similar levels on the bremsstrahlung rate and on the radius of the final star. It requires a direct photon coupling that cancels the one induced by ALP-pion mixing, and existing supernova gamma-ray limits may already cover it.
\end{abstract}

\maketitle

%=====================================================================
\section{Introduction}
%=====================================================================

Core-collapse supernovae are powerful laboratories to probe the emission of light particles with masses up to $\mathcal{O}(100)$ MeV \cite{Lella2023}; for reviews see Refs.~\cite{CaputoRaffelt2024rev,Carenza2025axastro}. The core of the protoneutron star (PNS) reaches temperatures of 20--50 MeV \cite{Pons1999}. Axion-like particles (ALPs) that couple to nucleons are produced there by nucleon-nucleon bremsstrahlung \cite{Carenza2019} and by pion conversion \cite{Lella2023}, which for light axions is 2--5 times larger than bremsstrahlung \cite{Carenza2021pion}. ALPs that leave the star carry away energy and shorten the neutrino burst, so the burst of SN 1987A limits their coupling to nucleons \cite{Lella2024}. Not every ALP leaves. The escape speed at the surface of the PNS is about $0.4c$ \cite{Bose2026ghosts}, and for an ALP with a mass within an order of magnitude of the temperature a non-negligible fraction of the emission is slower than this; it stays bound to the PNS and to the neutron star (NS) that the PNS becomes \cite{Lella2023,Bose2026ghosts}.

Bound populations of this kind were first pointed out for Kaluza-Klein gravitons from supernovae \cite{Lasenby2021,Hannestad2002}, to which we return below, and have been studied most closely around the Sun. Van Tilburg called them stellar basins and showed that, for a long-lived particle with a mass near the temperature of the solar core, the basin around the Sun outgrows the relativistic flux at the Earth within about a million years \cite{VanTilburg2021}. The solar basin has since been studied for dark photons \cite{Lasenby2021}, its orbits have been followed over the age of the Solar System with secular perturbation theory and direct integration \cite{Giovanetti2024}, and the photons from its decay have been searched for in X-ray data from NuSTAR \cite{DeRocco2022basin} and XMM-Newton \cite{Zhang2026basin}. Around pulsars, axions produced in the polar-cap region can also be gravitationally confined to the NS and accumulate into a dense cloud \cite{Noordhuis2024clouds}. A supernova leaves a basin of heavy ALPs around the NS, and this paper asks how much of it survives the star that it orbits.

Most work on supernova ALPs concerns the ones that escape. Heavy ALPs decay to photons on their way to the Earth, and the absence of these photons constrains their couplings. The non-detection by Fermi-LAT of decay photons from SN 1987A and Cas A constrains heavy ALPs produced by nucleon bremsstrahlung \cite{Giannotti2011}, and Fermi-LAT data towards Cas A, SN 1987A, SN 2023ixf and SN 2024ggi constrain heavy axions that couple to both photons and gluons through their time-delayed decay photons \cite{Benabou2025}. For ALPs coupled only to photons, the Gamma-Ray Spectrometer of the Solar Maximum Mission saw no excess after SN 1987A, which excludes a band of $g_{a\gamma\gamma}$ for $m_a\simeq10$ keV--100 MeV \cite{Jaeckel2018}; timing information changes this bound only mildly, because the gamma rays from early decays are absorbed close to SN 1987A \cite{Hoof2023}, and photon coalescence extends it to $m_a\approx280$ MeV \cite{Muller2023}. For photon-coupled ALPs, low-energy supernovae limit the energy that the decays deposit in the stellar envelope to about $0.1$ B, which excludes $g_{a\gamma\gamma}\sim10^{-10}$--$10^{-8}$ GeV$^{-1}$ for $m_a=1$--500 MeV \cite{Caputo2022lowE}. These bounds rest on the ALPs that escape. References~\cite{Lucente2020,Caputo2022lowE,Muller2023} remove the bound ones with an energy threshold for escape, which weakens the SN 1987A energy-loss bound on $g_{a\gamma\gamma}$ by at least 15\% for $m_a\gtrsim200$ MeV \cite{Lucente2020}.

The bound particles give a different signal. The idea that a supernova leaves the NS inside a halo of decaying particles goes back to Kaluza-Klein gravitons: their decay photons from nearby NSs were constrained with EGRET, and their decays would heat old NSs \cite{Hannestad2002,Hannestad2003}. For bulk majorons, which decay only to neutrinos, the neutrinos that hit the NS surface would heat it \cite{Hannestad2002majoron}. Lella \textit{et al.} considered the ALPs trapped around the NS in the Cas A remnant and the gamma rays from their decay \cite{Lella2023}, and Pospelov and Roychowdhury applied the idea to freeze-in dark matter that annihilates long after the supernova and heats the star \cite{Pospelov2026}.

Heating needs a cold star. A passively cooling NS reaches surface temperatures below $10^4$ K in less than $10^7$ yr \cite{Rodriguez2026heating}, so its thermal emission probes heat sources such as captured dark matter \cite{Baryakhtar2017}; for a review see Ref.~\cite{Bramante2024}. Internal heating, for example rotochemical heating, competes with these sources, and a very cold NS constrains the heating by dark matter \cite{Hamaguchi2019}. The coldest known NS is PSR J2144$-$3933, a pulsar with a characteristic age of 333 Myr that the Hubble Space Telescope did not detect; for a blackbody spectrum its surface temperature is below $4.2\times10^4$ K under conservative assumptions and for a range of NS radii and distances \cite{Guillot2019}. Other old pulsars are warmer. Hubble Space Telescope observations show likely thermal ultraviolet emission from PSR B0950+08, with a temperature far above passive cooling \cite{Abramkin2022}; for PSR J0108$-$1431 the far-ultraviolet flux, if it is thermal, implies a temperature that also requires heating \cite{Abramkin2021}. No single heating mechanism accounts for all of these stars together with the J2144$-$3933 limit \cite{Rodriguez2026heating}. Rotochemical heating is consistent with the J2144$-$3933 limit if the pulsar was born with a period of at least 10 ms \cite{Yanagi2020}.

Bose, Ganguly, Kumar, and Raj proposed that the bound supernova ALPs heat the old NS: the ALPs decay to two photons over the age of the star, and the photons that reach the surface are re-radiated as thermal emission \cite{Bose2026ghosts}. For the distance (192 pc), extinction ($E(B-V)=0.06$) and radius (11.9 km) that they adopt, the bounds on J2144$-$3933 are $3.4\times10^4$ K and a surface luminosity of $1.4\times10^{27}$ erg s$^{-1}$ [Ref.~\cite{Bose2026ghosts}, Eqs.~(11) and (12)]. With this star they exclude wide windows in the ALP-photon coupling $g_{a\gamma\gamma}$ for $m_a=1$--400 MeV, and the other NSs of their Table I, in particular PSR B0950+08 and the NS in Cas A, cover most of the remaining windows \cite{Bose2026ghosts}.

Reference~\cite{Bose2026ghosts} neglects the absorption of the ALPs by the star, because the ALPs are in the free-streaming regime. This is true for a single passage through the star. It is not true for the bound population, because a bound ALP whose orbit enters the star passes through it on every orbit. An orbit that reaches out to 50 km has a period of \PorbMs\ ms (Sec.~\ref{sec:cold}), so over 333 Myr the ALP crosses the star about \Ncross\ times.

The absorption of a bound population has been considered before. Van Tilburg noted that the basin an isolated NS builds around itself could be re-absorbed by the star as it cools, and that part of the basin could be protected on long-lived orbits outside the star \cite{VanTilburg2021}. For Kaluza-Klein gravitons, which couple with gravitational strength, Hannestad and Raffelt computed the back-absorption by inverse nucleon bremsstrahlung and found absorption times of 0.26 Gyr for one extra dimension and longer for more; in a 2025 revision they note that pionic absorption may dominate over inverse bremsstrahlung \cite{Hannestad2003}. Hannestad, Ker\"anen and Sannino estimated that bulk majorons are absorbed only on times much longer than the age of the old NS they considered \cite{Hannestad2002majoron}. The transfer of energy by feebly interacting bosons, including their reabsorption by inverse bremsstrahlung, has been formulated from free streaming to strong trapping, but without the gravitational redshift and the bending of trajectories that bound ALPs require \cite{Caputo2022transfer}. For photon-coupled axions in the trapping regime, Fiorillo \textit{et al.} note that a fully consistent treatment of gravity would be considerably more complex, because gravity bends the trajectories about as much as it redshifts the axions \cite{Fiorillo2025trapping}. An ALP with $g_{ap}=2.7\times10^{-10}$ GeV$^{-1}$, the value that Ref.~\cite{Bose2026ghosts} takes to satisfy the SN 1987A cooling bound, couples far more strongly to nucleons than Kaluza-Klein gravitons or bulk majorons, and its absorption by the cold star has not been calculated.

In this paper we calculate it. We compute the absorption of the bound ALPs in the cold star and in the PNS, follow every bound ALP along its general-relativistic orbit through the contracting PNS, and recompute the limits for the seven NSs of Ref.~\cite{Bose2026ghosts}. All the inputs are our own: twelve PNS evolutions with a finite-temperature equation of state, 19 equations of state for the cold star, and three bremsstrahlung rates, so that the spread of our results measures the uncertainty of the limits. Section~\ref{sec:formalism} collects the formalism, Secs.~\ref{sec:cold} and \ref{sec:surv} give the absorption in the cold star and the surviving population, Sec.~\ref{sec:limits} gives the limits, and Sec.~\ref{sec:disc} discusses what could still remove them.

Our results are:
(i) In the cold NS, inverse bremsstrahlung absorbs a bound ALP on a radial star-crossing orbit with apoapsis 50 km within \TabsRange\ at $m_a=10$--200 MeV for the 19 equations of state and three rates that we use, while for the couplings of the surviving windows the ALP decays in \TauMin\ s or longer.
(ii) Only orbits whose periapsis stays outside the final star survive, and they carry \RLrange\ of the heating.
(iii) The $np$ channel, which Ref.~\cite{Bose2026ghosts} includes in its numerical results but does not write out [N.~Raj, private communication], shortens the absorption time in the cold star by a factor \npColdHi\ at 100--200 MeV. At a fixed light-ALP energy loss of the supernova, the three rates give J2144$-$3933 margins within a factor \rateSpread\ of each other.
(iv) For PSR J2144$-$3933 a window survives absorption at \JmRange\ MeV for every one of our PNS evolutions, with its lower edge raised by a factor \JedgeRise\ at the masses we compute. PSR B0950+08, PSR J0108$-$1431, PSR B1727$-$47 and PSR J1836+5925 keep windows as well, PSR J0633+1746 keeps one only for the most favorable evolutions, and the NS in Cas A keeps none. With pion conversion and the $\Delta$ resonance, which raise the absorption above $m_\pi$ but the production of survivors more, the J2144$-$3933 window extends to \JmRangePi\ MeV.
(v) The surviving heating depends steeply on the radius $R$ of the final star: the J2144$-$3933 margin changes from \RsensHi\ to \RsensLo\ between $R=\RsensRlo$ and \RsensRhi\ km. In the two $\Delta$-admixed evolutions that we computed, $\Delta(1232)$ isobars change the margin by a factor \DeltaFac, mainly through the radius.
(vi) The surviving windows lie far below the photon coupling that ALP-pion mixing induces for a proton-coupled ALP \cite{Bauer2017,Benabou2025}, so they require a cancellation, and the supernova gamma-ray limits of Refs.~\cite{Lella2023,Benabou2025} have to be recast to decide whether they are open.

%=====================================================================
\section{Formalism}
\label{sec:formalism}
%=====================================================================

In this section we collect the equations that we solve, in the order in which the calculation uses them: the couplings and the absorption rates (Secs.~\ref{sec:formal_coup} and \ref{sec:formal_brems}), the dense matter and the PNS evolution that supply the densities and temperatures (Secs.~\ref{sec:formal_rmf} and \ref{sec:formal_pns}), and the bound population and its heating (Secs.~\ref{sec:formal_bound} and \ref{sec:formal_limits}). Sections~\ref{sec:cold}--\ref{sec:limits} then give the results.

\subsection{Couplings and decay}
\label{sec:formal_coup}

Reference~\cite{Bose2026ghosts} uses the ALP-hadron Lagrangian of heavy-baryon chiral perturbation theory (HBChPT) \cite{Ho2023}, whose nucleon and contact terms also appear in Ref.~\cite{Choi2022},
\begin{align}
\mathcal{L}_{\rm had}=-\partial_\mu a\Big[&g_{ap}\bar p\gamma^\mu\gamma_5p+g_{an}\bar n\gamma^\mu\gamma_5n\nonumber\\
&+\frac{ig_{a\pi N}}{f_\pi}\left(\pi^+\bar p\gamma^\mu n-\pi^-\bar n\gamma^\mu p\right)\nonumber\\
&+g_{aN\Delta}\left(\bar p\Delta^{+\mu}+\bar n\Delta^{0\mu}+{\rm h.c.}\right)\Big],
\label{eq:Lhad}
\end{align}
with $g_{a\pi N}=(g_{ap}-g_{an})/(\sqrt2\,g_A)$ and $g_{aN\Delta}=-\frac{\sqrt3}{2}\,(g_{ap}-g_{an})$ \cite{Ho2023}, where $f_\pi=92.4$ MeV is the pion decay constant and $g_A$ the axial coupling of the nucleon \cite{Bose2026ghosts}. We take, as Ref.~\cite{Bose2026ghosts} does, $g_{ap}=2.7\times10^{-10}$ GeV$^{-1}$ and $g_{an}=0$. With $g_A=1.28$, as in Ref.~\cite{Bose2026ghosts}, the contact and $\Delta$ couplings are then $g_{a\pi N}\simeq1.5\times10^{-10}$ GeV$^{-1}$ and $g_{aN\Delta}\simeq-2.3\times10^{-10}$ GeV$^{-1}$. For nucleon bremsstrahlung we keep only the first term. The contact term adds to the $np$ rate a piece of higher order in $1/m_N$, which is negligible \cite{Choi2022,Lella2023}. The other two terms change pion conversion, $\pi^-p\to na$, and its inverse, $an\to\pi^-p$, which arise already from the first term \cite{Lella2023,Lella2024,Ho2023}; we treat these processes at the end of Sec.~\ref{sec:formal_brems}. In the convention
\begin{equation}
\mathcal{L}_{aN}=\sum_{i=n,p}\frac{g_i}{2m_N}\,\bar\psi_i\gamma^\mu\gamma_5\psi_i\,\partial_\mu a
\label{eq:LaN}
\end{equation}
of Refs.~\cite{Giannotti2005,Carenza2019} this is $g_p=2m_Ng_{ap}=5.1\times10^{-10}$ and $g_n=0$, with $m_N$ the vacuum nucleon mass. Lella \textit{et al.} write the nucleon coupling in the form of Eq.~(\ref{eq:LaN}) and quote their limits for this dimensionless coupling \cite{Lella2023,Lella2024}, written as $C_pm_N/f_a$ in Ref.~\cite{Lella2024QCD}; with $g_{ap}=C_p/(2f_a)$ \cite{Bose2026ghosts} it equals $2m_Ng_{ap}$, and we convert their values to GeV$^{-1}$ with $m_N=0.939$ GeV. The photon coupling $\mathcal{L}_{a\gamma}=-\frac14 g_{a\gamma\gamma}\,aF_{\mu\nu}\tilde F^{\mu\nu}$ gives the decay rate \cite{Raffelt2007}
\begin{equation}
\Gamma_{a\to\gamma\gamma}=\tau_a^{-1}=\frac{g_{a\gamma\gamma}^2m_a^3}{64\pi},
\label{eq:decay}
\end{equation}
with $\tau_a$ the ALP lifetime. Below $3m_\pi$ this is the dominant decay, because the two-body decays $a\to\pi\pi$ and $a\to\pi^0\gamma$ are forbidden by CP and angular-momentum conservation and the three-body decays are suppressed by phase space and powers of the fine-structure constant $\alpha$ \cite{Bauer2017}; above $3m_\pi$ the decay $a\to3\pi$ opens \cite{Bauer2017,Bose2026ghosts}. For the photon couplings considered here, $g_{a\gamma\gamma}\lesssim10^{-13}$ GeV$^{-1}$, Primakoff production and conversion are negligible compared with the nucleon processes \cite{Bose2026ghosts,Lella2023}, so $g_{a\gamma\gamma}$ enters only through the decay.

A coupling to protons and not to neutrons has an isovector part, $(g_{ap}-g_{an})/2$, and below the QCD scale this part mixes the ALP with the neutral pion. The mixing induces a photon coupling through the $\pi^0\to\gamma\gamma$ anomaly \cite{Bauer2017,Benabou2025}. In the normalization of Ref.~\cite{Bauer2017} [their Eqs.~(10), (12), (20)--(22)], a coupling $c_u-c_d$ to light quarks gives $g_{a\gamma\gamma}^{\rm ind}=\alpha(c_u-c_d)m_a^2/[2\pi\Lambda(m_\pi^2-m_a^2)]$, with $\Lambda$ the scale of the ALP couplings. The same mixing adds to the proton coupling, so that $2\Lambda g_{ap}=g_A(c_u-c_d)\,m_\pi^2/(m_\pi^2-m_a^2)$ for $g_{an}=0$, as in Eq.~(29) of Ref.~\cite{Benabou2025}. At fixed physical $g_{ap}$ the pole cancels, and
\begin{equation}
|g_{a\gamma\gamma}^{\rm ind}|\simeq\frac{\alpha\,g_{ap}}{\pi g_A}\,\frac{m_a^2}{m_\pi^2}\simeq4.9\times10^{-13}\ {\rm GeV}^{-1}\,\frac{m_a^2}{m_\pi^2},
\label{eq:gind}
\end{equation}
for an ALP coupled only to light quarks (the number is for $g_A=1.28$), up to corrections of order $m_{u,d}/m_s$ from the strange quark \cite{Lella2024QCD}; a coupling to gluons adds a term that does not vanish for $m_a\to0$ \cite{Lella2024QCD}. This derivation is ours. For $m_a\ll m_\pi$ Eq.~(10) of Ref.~\cite{Lella2024QCD} gives the same coupling; it relates the proton coupling to the quark couplings without the pion pole, so it keeps the factor $m_\pi^2/(m_\pi^2-m_a^2)$. As a check, for $m_a$ close to $m_\pi$ the ALP couples through its $\pi^0$ admixture, and the ratio of Eq.~(\ref{eq:gind}) to the isovector coupling $g_{ap}/2$ approaches $2\alpha/(\pi g_A)$, the ratio of the $\pi^0$ couplings $\alpha/(\pi f_\pi)$ to photons and $g_A/(2f_\pi)$ to the proton axial current. Equation~(26) of Ref.~\cite{Benabou2025} as printed has a factor $\frac12$ in the quark term that Eq.~(7) of Ref.~\cite{Lella2024QCD} and Eq.~(22) of Ref.~\cite{Bauer2017} do not have, and gives half of Eq.~(\ref{eq:gind}). Equation~(\ref{eq:gind}) gives \gindTen\ GeV$^{-1}$ at 10 MeV and \gindHundred\ GeV$^{-1}$ at 100 MeV. A total photon coupling far below Eq.~(\ref{eq:gind}) needs a direct coupling that cancels it; without such tuning the induced coupling is present \cite{Lella2024QCD}. We treat $g_{a\gamma\gamma}$ as a free parameter, as Ref.~\cite{Bose2026ghosts} does, and return to this point in Sec.~\ref{sec:disc}.

The ALP needs no term in the Lagrangian of dense matter, Eq.~(\ref{eq:LRMF}) below. In the nonrelativistic limit it couples to the nucleon spin through $\nabla a$, as Bai \textit{et al.} note for neutrons \cite{Bai2025}, and in unpolarized matter the mean spin density that would act as its source vanishes; the mean fields therefore do not depend on $a$, and the ALP enters only through the rates.

\subsection{Absorption and emission by nucleon bremsstrahlung}
\label{sec:formal_brems}

Since the ALP enters only through the rates, we now give them, first for bremsstrahlung and then, at the end of this subsection, for pion conversion. An ALP of energy $\omega$ and momentum $\mathbf a$, with speed $\zeta=|\mathbf a|/\omega$, is absorbed in nucleon matter by $aNN\to NN$. The rate is
\begin{align}
\Gamma_{\rm abs}=\frac{1}{2\omega}\int&\prod_{i=1}^{4}\frac{d^3p_i}{(2\pi)^32E_i}\nonumber\\
&\times f_1f_2(1-f_3)(1-f_4)\,S\sum_{\rm spins}|\mathcal M|^2\nonumber\\
&\times(2\pi)^4\delta^4(p_1+p_2+a-p_3-p_4),
\label{eq:gamma}
\end{align}
where $p_i=(E_i,\mathbf p_i)$ and $a$ are the four-momenta of the nucleons and the ALP, the $f_i$ are nucleon occupation numbers, the sum runs over all nucleon spins, and $S$ is the symmetry factor ($S=1/4$ for $pp$, $S=1$ for $np$). For nonrelativistic nucleons of mass $m$, $E_i\simeq m+p_i^2/2m$, the momentum and energy delta functions remove the final momenta, and
\begin{align}
\Gamma_{\rm abs}=\;&\frac{1}{64\,\omega\,(2\pi)^8m^3}\int d^3p_1\,d^3p_2\,d\Omega_{\hat q}\nonumber\\
&\times q\,f_1f_2(1-f_3)(1-f_4)\,S\sum|\mathcal M|^2,
\label{eq:gammaC}
\end{align}
with $\mathbf p_{3,4}=\mathbf P\pm q\hat{\mathbf q}$, where $\mathbf P=(\mathbf p_1+\mathbf p_2+\mathbf a)/2$ and $q^2=(p_1^2+p_2^2)/2+m\omega-P^2$.

We take the matrix elements from one-pion exchange (OPE) \cite{Brinkmann1988,Giannotti2005}. With $\mathbf k=\mathbf p_2-\mathbf p_4$, $\mathbf l=\mathbf p_2-\mathbf p_3$, $F_k=k^2/(k^2+m_\pi^2)$, $F_l=l^2/(l^2+m_\pi^2)$, $\xi=3(\hat{\mathbf k}\cdot\hat{\mathbf l})^2$ and the pion-nucleon coupling constant $f\simeq1$, the $pp$ channel is \cite{Giannotti2005}
\begin{equation}
S\sum|\mathcal M|^2_{pp}=\frac{64}{3}\frac{\zeta^2f^4m^2g_p^2}{m_\pi^4}\left[F_k^2+F_l^2+(1-\xi)F_kF_l\right],
\label{eq:Mpp}
\end{equation}
and the $np$ channel, with nucleons 1, 3 neutrons and 2, 4 protons, is
\begin{align}
\sum|\mathcal M|^2_{np}=\;&\frac{128}{3}\frac{\zeta^2f^4m^2}{m_\pi^4}\Big[(g_n^2+g_p^2)F_k^2\nonumber\\
&+(3g_n^2+3g_p^2+2g_ng_p)F_l^2-\Big(2(g_n^2+g_p^2)\nonumber\\
&-\tfrac43(g_n^2+g_p^2+g_ng_p)\,\xi\Big)F_kF_l\Big].
\label{eq:Mnp}
\end{align}
The factor $\zeta^2$ is the velocity factor of a derivative coupling for a massive ALP \cite{Giannotti2005,Lella2023}. The sign of the interference term differs in the literature. References~\cite{Carenza2019,Choi2022} find the minus sign of Eq.~(\ref{eq:Mnp}), from the exchange of the neutron and the proton. Reference~\cite{Giannotti2005} has the opposite sign [their Eq.~(23) and footnote 14]; the coefficient $\frac43$ follows from their Eqs.~(19), (22) and (30), while their Eq.~(23) has $\frac43|\mathbf k\cdot\mathbf l|^2=\frac49\xi k^2l^2$, a factor 3 smaller. We have settled the sign by an explicit sum over the spins and isospins of the antisymmetrized two-nucleon states, with the static one-pion-exchange potential $-(f/m_\pi)^2(\boldsymbol\sigma_1\cdot\mathbf q)(\boldsymbol\sigma_2\cdot\mathbf q)\,\boldsymbol\tau_1\cdot\boldsymbol\tau_2/(q^2+m_\pi^2)$ and the amplitude $[H_a,V]/\omega$ of emission from all four legs, where $H_a=\sum_i(g_i/2m)\boldsymbol\sigma_i\cdot\mathbf a$. At random momenta it reproduces Eq.~(\ref{eq:Mpp}) and Eq.~(\ref{eq:Mnp}) with the minus sign, including the relative normalization of the two channels and the $g_ng_p$ term, to machine precision, and it is incompatible with the plus sign. All our rates use the minus sign. With $g_n=0$ the $np$ channel contributes because the ALP is emitted from the proton line.

OPE overestimates the rate \cite{Carenza2019,Hanhart2001}; for a review see Ref.~\cite{Carenza2025axastro}. Following Ref.~\cite{Carenza2019}, we replace $(f/m_\pi)^4$ by $(g_A/2f_\pi)^4$ with $g_A=1.26$ and $f_\pi=92.4$ MeV [their Eq.~(2.9)], and, as proposed in Ref.~\cite{Ericson1989}, represent two-pion exchange by $\rho$ exchange [Ref.~\cite{Carenza2019}, Eqs.~(2.19) and (5.20)],
\begin{equation}
F_k\to\frac{k^2}{k^2+m_\pi^2}-C_\rho\frac{k^2}{k^2+m_\rho^2},
\label{eq:rho}
\end{equation}
with $C_\rho=1.67$ and $m_\rho=600$ MeV, and likewise for $F_l$. We do not include the multiple-scattering correction of Ref.~\cite{Carenza2019}, which can lower the OPE rate by more than half above $1.5\times10^{14}$ g cm$^{-3}$ [their Fig.~8] but changes the rate little once the $\rho$ exchange and the other corrections are included [their Figs.~10 and 11]; we return to it in Sec.~\ref{sec:disc}.

We use three rates. Our baseline is this model with the $pp$ and $np$ channels. The second is its $pp$ channel alone, the process and $\rho$-exchange model written out in Ref.~\cite{Bose2026ghosts} [their Eq.~(5)]; their numerical results also include the $np$ channel [N.~Raj, private communication]. Their rate takes the nucleon mass from the supernova profile where we use the vacuum mass. The third is OPE with $pp$ and $np$, without the $\rho$ term.

In the hot PNS the nucleons have nonrelativistic Fermi-Dirac distributions $f_i=[e^{(p^2/2m-\mu_i)/T}+1]^{-1}$ with the vacuum mass, and the kinetic chemical potentials $\mu_i$ are fixed by the densities $n_i$; $T$ is the temperature. These are not the nucleon distributions of the relativistic mean-field (RMF) matter of Eq.~(\ref{eq:LRMF}), in which the effective mass $M^*$ is lower and the nucleons are more degenerate: at equal densities the degeneracy parameter $(\nu_i-M^*)/T$ of the RMF, with $\nu_i$ the effective chemical potential defined below Eq.~(\ref{eq:LRMF}), exceeds that of the vacuum-mass distribution by \EtaGapGM\ for GM3 and \EtaGapFS\ for FSU2R, at four states with baryon densities $n_B=0.1$--0.45 fm$^{-3}$ and $T=10$--30 MeV. Both the lower mass \cite{Carenza2019} and the stronger Pauli blocking lower the rate, so in this respect our hot absorption is an overestimate. In the cold star $f_i=\theta(p_{F,i}-p)$, with $p_{F,i}$ the Fermi momenta. For the OPE rate in the cold star we use the Landau effective mass $m=(p_{F,n}^2+M^{*2})^{1/2}$ of the RMF model in the kinematics and the vacuum mass in the vertex, so that $|\mathcal M|^2$ scales as $m^4/m_N^2$, and we add the nucleon-velocity term of Bai \textit{et al.} \cite{Bai2025} [their Eqs.~(3.9) and (S14)], derived for neutron matter, which replaces $\zeta^2$ by $\zeta^2+\frac32p_F^2/m^2$; we take $p_F$ as the proton Fermi momentum. We do not include this term in the $\rho$-exchange rates or in the hot PNS, where it would raise the absorption of slow ALPs; the survival fractions below are optimistic in this respect. For the two $\rho$-exchange rates we use the vacuum mass. An effective mass $m=0.6$--$0.9\,m_N$ in the kinematics lowers the $\rho$-exchange rate by a factor 0.36--0.93 in our tests, in line with the reduction of Ref.~\cite{Carenza2019} [their Fig.~7].

The emission rate into the same mode follows from detailed balance,
\begin{equation}
\Gamma_{\rm em}(\omega)=\Gamma_{\rm abs}(\omega)\,e^{-\omega/T},
\label{eq:db}
\end{equation}
so a single function gives both the production and the absorption. The number of ALPs produced per unit volume, time and speed is
\begin{equation}
\frac{dn}{dv\,d\tau}=\frac{m_a^3\gamma^5v^2}{2\pi^2}\,\Gamma_{\rm abs}(v)\,e^{-\gamma m_a/T},
\label{eq:prod}
\end{equation}
with $v=\zeta$ the ALP speed, $\gamma=(1-v^2)^{-1/2}$, or $d^2n/dE\,dt=pE\,\Gamma_{\rm abs}e^{-E/T}/2\pi^2$ in terms of the energy $E=\gamma m_a$ and momentum $p=\gamma m_av$. This is the spectrum of Eq.~(5) of Ref.~\cite{Bose2026ghosts}.

We evaluate Eq.~(\ref{eq:gammaC}) as an eight-dimensional integral with the adaptive Monte Carlo method VEGAS \cite{Lepage1978,Lepage2021}. Three checks fix the normalization. In nondegenerate matter the integral agrees with the two-dimensional expression of Ref.~\cite{Giannotti2005} to 0.2\% at $T=10$--30 MeV. In degenerate matter, for $\omega$ well below the Fermi energy and with the $\mathbf k\cdot\mathbf l$ term dropped as in Ref.~\cite{Bai2025}, it agrees with their Fermi-surface result to 2--5\%. At finite temperature the ratio of absorption to emission equals $e^{\omega/T}$ to $5\times10^{-4}$. For the PNS we tabulate $\Gamma_{\rm abs}$ at $\zeta=0.5$ on grids in $n_B$ (or the proton density), $T$, the proton fraction $Y_p$ and $m_a$. For the $\rho$-exchange rates we use $\Gamma_{\rm abs}(\zeta)=\Gamma_{\rm abs}(0.5)\,(\zeta/0.5)^2/[1+0.75(\zeta^2-0.25)]$, which reproduces direct evaluations of the $np$ channel to 13\%, while the pure $\zeta^2$ scaling stays within 15\% of the $pp$ channel for $\zeta\le0.5$; for OPE we keep $(\zeta/0.5)^2$. Speeds above 0.95 are set to 0.95; they do not occur for bound ALPs.

\textit{Pion conversion and the $\Delta$ resonance.} For $\omega>m_\pi$ an ALP can also be absorbed by $aN\to\pi N$ and produced by the inverse process. For massless QCD axions, Carenza \textit{et al.} found pion conversion 2--5 times larger than nucleon bremsstrahlung, with a harder spectrum \cite{Carenza2021pion}. For massive ALPs the contact and $\Delta$ terms of Eq.~(\ref{eq:Lhad}) each raise the pion-conversion emission by at least a factor 2 for $m_a\lesssim500$ MeV \cite{Lella2023}, and for the couplings of the Kim-Shifman-Vainshtein-Zakharov (KSVZ) and Dine-Fischler-Srednicki-Zhitnitsky (DFSZ) axion models the $\Delta$ term interferes destructively above $T\approx55$ MeV \cite{Ho2023}. These terms change the production and the absorption together. We include the channels $an\leftrightarrow\pi^-p$, $ap\leftrightarrow\pi^+n$, $ap\leftrightarrow\pi^0p$ and $an\leftrightarrow\pi^0n$ with the HBChPT amplitude of Ref.~\cite{Ho2023}, which contains the nucleon-pole, contact and $\Delta$-mediated diagrams of Eq.~(\ref{eq:Lhad}). The direct $\Delta$-mediated diagram is the one-body absorption $aN\to\Delta$ followed by $\Delta\to\pi N$; the crossed one is not resonant. For a proton at rest Ref.~\cite{Ho2023} finds $\sigma(\pi^-p\to na)=E_\pi m^2G_a/(16\pi f_\pi^2f_a^2|\mathbf k_\pi|)$, with $E_\pi$ and $\mathbf k_\pi$ the pion energy and momentum and $f_a$ the axion decay constant. We use the leading-order part of $G_a$ of their Eq.~(39), as in Ref.~\cite{Lella2023}, the $\Delta$-nucleon mass difference $\Delta m=293$ MeV, the $\Delta$ width $\Gamma_\Delta=117$ MeV and the $\pi N\Delta$ coupling $3g_A/2$; in our notation $C_x/f_a\to2g_x$. For a massive ALP the squared amplitude carries an extra factor $\zeta^2$ \cite{Lella2023}. With static nucleons, so that $E_\pi=\omega$, the absorption rate on a nucleon $N_1$ that turns into $N_2$ is
\begin{equation}
\Gamma_{\rm abs}^{(\pi)}=\eta_{12}\,|\mathbf k_\pi|\,\big[1+n_\pi(\omega)\big]\,\frac{\zeta^2m^2G_a}{8\pi\,\omega f_\pi^2},
\label{eq:gpion}
\end{equation}
where $\eta_{12}=\int d^3p\,f_1(1-f_2)/(2\pi)^3$ for one spin state and $n_\pi$ is the Bose-Einstein distribution of the pions, with $\mu_{\pi^\mp}=\pm(\mu_n-\mu_p)$ and $\mu_{\pi^0}=0$. Equation~(\ref{eq:gpion}) reduces to $n_N\sigma v_a$ for a nucleon at rest, with $n_N$ the nucleon density and $v_a=\zeta$, and with the same nucleon distributions as the bremsstrahlung rates it obeys Eq.~(\ref{eq:db}) channel by channel. For $\pi^+n$ we exchange $g_{ap}\leftrightarrow g_{an}$ in $G_a$ \cite{Lella2024}, which reverses the isoscalar part of the nucleon-pole--$\Delta$ interference. The interference of the contact term with the other diagrams, whose sign differs between the two charged channels, is of higher order in $1/m$ and is not included \cite{Ho2023,Lella2024}. The $\pi^0$ channels have no contact term. Their nucleon-pole term has the same nucleon at both vertices and half the squared pion coupling of the charged channels \cite{Lella2024}. By the isospin Clebsch-Gordan coefficients their $\Delta$ term is twice as large, with the same sign for the direct and crossed diagrams, which have opposite signs in the charged channels; Ref.~\cite{Lella2024} omits this term. We keep the interference of the two.

We use vacuum pions. The absorption depends on the pion abundance only through the Bose factor $1+n_\pi$ \cite{Lella2024}, but the production by $\pi^-p\to na$ grows with the $\pi^-$ abundance, which in-medium interactions raise \cite{Fore2020}, so our $\pi^-p$ production is likely a lower estimate. At $\omega=300$--400 MeV the $\pi^0$ channels, which proceed mainly through the $\Delta$ and which Ref.~\cite{Bose2026ghosts} does not include, give \PiZeroShare\% of the pion-conversion production. In our PNS matter the absorption by Eq.~(\ref{eq:gpion}) exceeds inverse bremsstrahlung by factors \PiAbsRatio\ at $m_a=200$--300 MeV, and in the $\pi^-p$ channel the $\Delta$ term alone gives \PiDeltaShare\% of $G_a$ at $\omega=300$ MeV. Within the $1/m$ expansion of Ref.~\cite{Ho2023}, the terms of next order, the third and last terms of their Eq.~(39) and $\Delta G_a$ of their Eq.~(43), lower our rate by \PiHoFac\% at $\omega=400$ MeV. This is a different comparison from that of Ho \textit{et al.}, who find that the cross section and the emission rate evaluated with the large-$m_N$ expansion exceed those evaluated without it [their Figs.~2--4], so that the expansion overestimates the emission \cite{Ho2023}. We keep the leading order, as Ref.~\cite{Lella2023} does; by Eq.~(\ref{eq:db}) an overestimate of the production is an overestimate of the absorption by the same factor. We show the limits with and without these channels.
\subsection{Dense matter at zero and finite temperature}
\label{sec:formal_rmf}

The rates depend on the nucleon densities, effective masses and temperature. For both the cold NS and the PNS we take these from RMF models of nucleons coupled to $\sigma$, $\omega$ and $\rho$ mesons,
\begin{align}
\mathcal{L}=&\sum_{i=n,p}\bar\psi_i\Big[\gamma^\mu\Big(i\partial_\mu-g_\omega\omega_\mu-\frac{g_\rho}{2}\bm\tau\cdot\bm\rho_\mu\Big)-M^*\Big]\psi_i\nonumber\\
&+\frac12\partial_\mu\sigma\partial^\mu\sigma-\frac12m_\sigma^2\sigma^2-U(\sigma)-\frac14\omega_{\mu\nu}\omega^{\mu\nu}\nonumber\\
&+\frac12m_\omega^2\omega_\mu\omega^\mu-\frac14\bm\rho_{\mu\nu}\cdot\bm\rho^{\mu\nu}+\frac12m_\rho^2\bm\rho_\mu\cdot\bm\rho^\mu\nonumber\\
&+\mathcal{L}_{\omega\rho}+\mathcal{L}_{\rm lep},
\label{eq:LRMF}
\end{align}
with $M^*=M-g_\sigma\sigma$, $U(\sigma)=\frac{bM}{3}(g_\sigma\sigma)^3+\frac{c}{4}(g_\sigma\sigma)^4$, and $\tau_3=+1$ for neutrons. Here $\psi_i$ are the nucleon fields, $M$ is the nucleon mass, $g_\sigma$, $g_\omega$, $g_\rho$ and $m_\sigma$, $m_\omega$, $m_\rho$ are the meson-nucleon couplings and the meson masses, $b$ and $c$ are the scalar self-couplings, $\omega_{\mu\nu}$ and $\bm\rho_{\mu\nu}$ are the vector-meson field tensors, and $\mathcal{L}_{\rm lep}$ describes free leptons. The term $\mathcal{L}_{\omega\rho}$ contains the nonlinear meson couplings ($\omega^4$, $\omega^2\rho^2$, $\sigma$-$\omega$ and $\sigma$-$\rho$) in the form and with the parameters of Ref.~\cite{Sun2024} [the functions $F$ and $G$ of their Eq.~(15)].

For the cold stars we use \NsetsPass\ RMF sets from the compilation of Ref.~\cite{Sun2024}. They are the sets that pass two observational tests in cold, $\beta$-equilibrated matter: a maximum mass of at least 2.01 $M_\odot$, the lower edge of the 68.3\% credible interval for the mass of PSR J0740+6620 \cite{Fonseca2021}, and a binary tidal deformability $\tilde\Lambda\le720$ for two 1.365 $M_\odot$ stars, the upper edge of the 90\% highest-posterior-density interval of GW170817 for low spins \cite{Abbott2019}. Of the \NsetsAll\ meson-exchange sets of Ref.~\cite{Sun2024} that we can solve through the maximum-mass star, \NsetsMass\ pass the first test and \NsetsPass\ both. For these tests we use the physical neutron and proton masses and the crust of Baym, Pethick and Sutherland \cite{Baym1971}; with a single nucleon mass and the SLy crust, which we use below for the cold stars, RMF423 gives $\tilde\Lambda=725$ and would narrowly fail the second test. These sets give radii of \RcoldRange\ km at 1.35 $M_\odot$. Of these, HA also contains a $\delta$ meson and TW-99 has density-dependent couplings with the rearrangement term; we include both as in Ref.~\cite{Sun2024}. For the PNS we use GM3 \cite{Glendenning1991}, with the parameters of Table 1 of Ref.~\cite{Pons1999}, and FSU2R \cite{Tolos2017}.

In the mean-field approximation the meson fields are replaced by their expectation values $\bar\sigma$, $\bar\omega$, $\bar\rho$. At temperature $T$ each nucleon has the effective mass $M^*=M-g_\sigma\bar\sigma$, the effective chemical potential $\nu_i=\mu_i-g_\omega\bar\omega\mp\frac12g_\rho\bar\rho$ (upper sign for neutrons), and the occupation $f_i(p)=[e^{(E^*-\nu_i)/T}+1]^{-1}$ with $E^*=(p^2+M^{*2})^{1/2}$. The vector and scalar densities are
\begin{equation}
n_i=\frac{1}{\pi^2}\int_0^\infty p^2f_i\,dp,\qquad n_{s,i}=\frac{1}{\pi^2}\int_0^\infty p^2\frac{M^*}{E^*}f_i\,dp,
\label{eq:dens}
\end{equation}
and the fields solve
\begin{align}
m_\sigma^2\bar\sigma+U'(\bar\sigma)&=g_\sigma(n_{s,n}+n_{s,p}),\nonumber\\
\frac{\partial\mathcal{U}_{\omega\rho}}{\partial\bar\omega}=g_\omega n_B,&\qquad
\frac{\partial\mathcal{U}_{\omega\rho}}{\partial\bar\rho}=\frac{g_\rho}{2}(n_n-n_p),
\label{eq:fields}
\end{align}
where $\mathcal{U}_{\omega\rho}=\frac12m_\omega^2\bar\omega^2+\frac12m_\rho^2\bar\rho^2+\mathcal{L}_{\omega\rho}$ \cite{Tolos2017}. For the sets of Ref.~\cite{Sun2024} that also couple $\sigma$ to $\omega$ and $\rho$, the $\sigma$ equation contains in addition the derivative of $\mathcal{L}_{\omega\rho}$ with respect to $\bar\sigma$, and we include it. The energy density and pressure follow from the Fermi-Dirac integrals and the field energies, and the entropy density from $Ts=\varepsilon+P-\sum_i\mu_in_i$ \cite{Prakash1997}. Antinucleons are negligible. We have checked this implementation in three ways. For symmetric matter at $T\to0$ it gives for GM3 the saturation density, energy per nucleon, incompressibility, symmetry energy and Dirac effective mass \GMsat, close to the values at which the model is calibrated, $n_0=0.153$ fm$^{-3}$, $E_0=-16.3$ MeV, $K_0=240$ MeV, $J=32.5$ MeV and $M^*/M=0.78$ [Ref.~\cite{Pons1999}, Table 1], and for the 19 sets of Ref.~\cite{Sun2024} it reproduces the tabulated saturation density, binding energy, incompressibility and symmetry energy to \SatDev\%. At finite temperature and fixed composition the pressure and the neutron chemical potential agree with numerical derivatives of the free energy to \ThermoDev\ at six states of GM3 and FSU2R with $n_B=0.1$--0.5 fm$^{-3}$ and $T=10$--30 MeV, and the solutions of Eq.~(\ref{eq:fields}) make the grand potential stationary in each field at fixed chemical potentials.

The PNS matter also contains electrons, positrons, trapped electron neutrinos, $\nu_\mu$, $\nu_\tau$ and their antiparticles, and photons. The leptons are massless where they matter, so for a species with degeneracy $g$ and chemical potential $\mu$ the net density and pressure of particles and antiparticles are
\begin{align}
n&=\frac{g}{6\pi^2}\left(\mu^3+\pi^2T^2\mu\right),\nonumber\\
P&=\frac{g}{24\pi^2}\left(\mu^4+2\pi^2T^2\mu^2+\frac{7\pi^4}{15}T^4\right),
\label{eq:lep}
\end{align}
with $g=2$ for electrons, $g=1$ for $\nu_e$, and $g=2$ and $\mu=0$ for $\nu_\mu$ and $\nu_\tau$ together; photons add $P_\gamma=\pi^2T^4/45$. Muons and pions are omitted. Thermal pions raise the proton fraction and soften the equation of state, and reactions with pions and muons add to the opacity of muon neutrinos \cite{Fore2020}; Fischer \textit{et al.} include pions in the equation of state of supernova simulations \cite{Fischer2021axion}. Beta equilibrium with trapped neutrinos and charge neutrality,
\begin{equation}
\mu_n=\mu_p+\mu_e-\mu_{\nu_e},\qquad n_p=n_{e^-}-n_{e^+},
\label{eq:beta}
\end{equation}
fix the composition at given $(n_B,T,\mu_{\nu_e})$, and the lepton fraction is $Y_L=Y_e+Y_{\nu_e}$ \cite{Prakash1997}. We tabulate all thermodynamic quantities on a grid in $(\ln n_B,\ln T,\mu_{\nu_e})$. For GM3 the maximum masses at $(Y_L=0.4,s=1)$, $(Y_\nu=0,s=2)$ and $(Y_\nu=0,s=0)$ are \PonsMgOurs\ $M_\odot$, within \PonsMgDev\% of Table 2 of Ref.~\cite{Pons1999} (1.946, 2.044 and 2.005 $M_\odot$); our tables have neither muons nor a crust.

\textit{$\Delta$ isobars.} Above saturation density the $\Delta(1232)$ resonances can appear in neutron-star matter for a wide range of meson-$\Delta$ couplings \cite{Li2018Delta}; they soften the equation of state at intermediate densities, stiffen it at high densities \cite{Li2018Delta,Sedrakian2022Delta}, and make stars more compact \cite{Li2018Delta,Thapa2021,Parmar2025}. We add the non-strange quartet $\Delta^-,\Delta^0,\Delta^+,\Delta^{++}$ as Rarita-Schwinger fields,
\begin{equation}
\mathcal{L}_\Delta=\sum_{d}\bar\psi_{d\nu}\Big[\gamma^\mu\Big(i\partial_\mu-x_\omega g_\omega\omega_\mu-x_\rho g_\rho\,t_{3d}\,\rho_{3\mu}\Big)-M_\Delta^*\Big]\psi_d^\nu,
\label{eq:LDelta}
\end{equation}
where $x_\sigma$, $x_\omega$ and $x_\rho$ are the ratios of the $\Delta$-meson to the nucleon-meson couplings, $M_\Delta^*=M_\Delta-x_\sigma g_\sigma\sigma$, $M_\Delta=1232$ MeV, and $t_{3d}=\frac32,\frac12,-\frac12,-\frac32$ in the isospin convention of Eq.~(\ref{eq:LRMF}) (neutron $t_3=+\frac12$). In mean field each $\Delta$ is a Fermi gas of degeneracy 4 with effective chemical potential $\nu_d=\mu_d-x_\omega g_\omega\bar\omega-x_\rho g_\rho t_{3d}\bar\rho$; the $\Delta$ densities enter the sources of Eq.~(\ref{eq:fields}) with the factors $x_\sigma$, $x_\omega$ and $x_\rho$, and chemical equilibrium gives $\mu_{\Delta^-}=2\mu_n-\mu_p$, $\mu_{\Delta^0}=\mu_n$, $\mu_{\Delta^+}=\mu_p$, $\mu_{\Delta^{++}}=2\mu_p-\mu_n$. With the $\Delta$s the pressure agrees with $n_B^2\partial(f/n_B)/\partial n_B$ to \DeltaThermo\ at $n_B=0.6$ fm$^{-3}$ and $T=0.5$ MeV in beta equilibrium (GM3, $x_\sigma=1.18$), and for $x_\sigma=0$ the $\Delta$s are absent and the nucleonic equation of state is recovered exactly.

The couplings are poorly known. Reference~\cite{Sedrakian2022Delta} takes all three ratios equal to 1; Ref.~\cite{Yogeesh2026} takes $x_\omega=x_\rho=1$ and $x_\sigma=1.18$; Ref.~\cite{Thapa2021} takes $x_\omega=1.10$, $x_\rho=1$ and $x_\sigma=1.10$ or 1.23; and the Bayesian analysis of Ref.~\cite{Parmar2025}, with $x_\rho=1$, finds $x_\sigma\simeq1.35$--1.38 and $x_\omega\simeq1.18$. All four works fix these ratios with density-dependent nucleon couplings, not with GM3 or FSU2R, so they guide our choice but do not fix it. We take $x_\omega=x_\rho=1$ and $x_\sigma=1.0$--1.18, which spans the first two choices and lies inside the range $x_\rho=1$, $0.8\le x_\omega\le1.6$, $x_\sigma=x_\omega\pm0.2$ quoted in Ref.~\cite{Sedrakian2022Delta}. A larger $x_\sigma$ lowers $M_\Delta^*$ relative to the nucleon mass and moves the onset to lower density. We neglect the $\Delta$ contributions to the neutrino opacities, and in the ALP rates we use the nucleon densities and the nucleon proton fraction.

The $\Delta$ also opens a one-body absorption channel, $aN\to\Delta$, through the $g_{aN\Delta}$ term of Eq.~(\ref{eq:Lhad}) \cite{Ho2023}. For a nucleon and a $\Delta$ both at rest, the threshold is the difference of their single-particle energies, $M_\Delta^*-M^*$ plus the difference of the vector potentials; for $x_\sigma=x_\omega=x_\rho=1$ the meson terms cancel for $p\to\Delta^+$ and $n\to\Delta^0$ and the threshold is $M_\Delta-M=293$ MeV. Because $E_\Delta(p)-E_N(p)$ falls with momentum, a nucleon at the Fermi surface needs less: about \DthFermi\ MeV at the center of the cold FSU2R star with $x_\sigma=1$ and of the GM3 star with $x_\sigma=1.18$. In the hot PNS we include this channel, with vacuum nucleons and $\Delta$s, through the $\Delta$-mediated diagram of pion conversion, Eq.~(\ref{eq:gpion}). In the cold star it changes nothing, since every orbit that enters the star is already absorbed.

\subsection{Protoneutron-star evolution}
\label{sec:formal_pns}

The hot rates are needed along the PNS profiles $n_B(r,t)$, $T(r,t)$ and $Y_p(r,t)$ of the first 10 s, which we compute with the finite-temperature equation of state above. We follow the quasi-static evolution of a PNS of fixed baryon number with the equilibrium diffusion approximation of Pons \textit{et al.} \cite{Pons1999}. In the metric
\begin{equation}
ds^2=-e^{2\phi}dt^2+e^{2\Lambda}dr^2+r^2d\Omega^2,\qquad e^{-2\Lambda}=1-\frac{2m}{r},
\label{eq:metric}
\end{equation}
($G=c=1$) the structure in the enclosed baryon number $a$ obeys [Ref.~\cite{Pons1999}, Eqs.~(34), (35)]
\begin{align}
\frac{\partial r}{\partial a}&=\frac{1}{4\pi r^2n_Be^{\Lambda}},\qquad\frac{\partial m}{\partial a}=\frac{\varepsilon}{n_Be^{\Lambda}},\nonumber\\
\frac{\partial\phi}{\partial a}&=\frac{e^{\Lambda}(m+4\pi r^3P)}{4\pi r^4n_B},\qquad\frac{\partial P}{\partial a}=-(\varepsilon+P)\frac{\partial\phi}{\partial a},
\label{eq:struct}
\end{align}
with $\phi$ matched to the Schwarzschild value $\frac12\ln(1-2M/R)$ at the surface. The neutrino number and energy transport equations [Ref.~\cite{Pons1999}, Eqs.~(13), (14)], combined with the evolution of the electron fraction and the matter energy equation [their Eqs.~(36), (38)] and with the first law of thermodynamics, give
\begin{align}
\frac{\partial Y_L}{\partial t}+\frac{\partial(e^{\phi}4\pi r^2F_\nu)}{\partial a}&=0,\label{eq:YL}\\
e^{\phi}\left(T\frac{\partial s}{\partial t}+\mu_\nu\frac{\partial Y_L}{\partial t}\right)+\frac{\partial(e^{2\phi}4\pi r^2H_\nu)}{\partial a}&=0,\label{eq:s}
\end{align}
where $s$ is the entropy per baryon. These are the conservative forms, which follow from Eqs.~(13), (14), (36) and (38) of Ref.~\cite{Pons1999}; Eqs.~(37) and (39) of Ref.~\cite{Pons1999} print an extra factor $e^{-\phi}$ in front of the flux divergence that does not follow from those equations, and with it our discretization loses \ConsLeptonLoss\% of the lepton number in the first second. With Eqs.~(\ref{eq:YL}) and (\ref{eq:s}) lepton number is conserved exactly, and the radiated energy equals the loss of gravitational mass to within \EconsRange\% over 10 s in all twelve evolutions. In the diffusion approximation the number and energy fluxes are [Ref.~\cite{Pons1999}, Eq.~(31)]
\begin{align}
F_\nu&=-\frac{e^{-\Lambda}e^{-\phi}T^2}{6\pi^2}\left[D_3\frac{\partial(Te^{\phi})}{\partial r}+(Te^{\phi})D_2\frac{\partial\eta}{\partial r}\right],\nonumber\\
H_\nu&=-\frac{e^{-\Lambda}e^{-\phi}T^3}{6\pi^2}\left[D_4\frac{\partial(Te^{\phi})}{\partial r}+(Te^{\phi})D_3\frac{\partial\eta}{\partial r}\right],
\label{eq:flux}
\end{align}
with $\eta=\mu_\nu/T$ and the diffusion coefficients
\begin{equation}
D_n=\int_0^\infty dx\,x^nD(\omega)f_0(1-f_0),\qquad x=\omega/T,
\label{eq:Dn}
\end{equation}
summed over species as $D_2=D_2^{\nu_e}+D_2^{\bar\nu_e}$, $D_3=D_3^{\nu_e}-D_3^{\bar\nu_e}$, $D_4=D_4^{\nu_e}+D_4^{\bar\nu_e}+4D_4^{\nu_\mu}$ [Ref.~\cite{Pons1999}, Eqs.~(32), (33)]. Here $f_0$ is the equilibrium neutrino distribution and $D(\omega)=1/[\kappa_a(\omega)+\kappa_s(\omega)]$ is the mean free path, with $\kappa_a$ the absorption opacity corrected for stimulated absorption and $\kappa_s$ the transport scattering opacity [Ref.~\cite{Pons1999}, Eq.~(28)].

For the opacities we use the free-nucleon cross sections of Burrows, Reddy and Thompson \cite{Burrows2004} with the in-medium kinematics and blocking of the RMF model. The absorption opacity for $\nu_en\to e^-p$ is
\begin{equation}
\kappa_a^{\nu_e}=\sigma_0\frac{1+3g_A^2}{4}\left(\frac{E_e}{m_e}\right)^2\frac{[1-f_e(E_e)]\,\eta_{np}}{1-f_0(\omega)},
\label{eq:kabs}
\end{equation}
where $E_e=\omega+Q$ is the electron energy (we drop the factor $[1-(m_e/E_e)^2]^{1/2}$ of Ref.~\cite{Burrows2004}, since $E_e\gg m_e$),
$\sigma_0=1.705\times10^{-44}$ cm$^2$, $g_A=1.26$ (Ref.~\cite{Burrows2004} uses 1.23), $Q=m_n-m_p+U_n-U_p$ the threshold shifted by the mean-field potentials, $1/(1-f_0)$ the stimulated-absorption correction \cite{Burrows2004}, and $\eta_{np}=(n_p-n_n)/\{\exp[(\nu_p-\nu_n)/T]-1\}$ the final-state blocking of the elastic approximation of Bruenn \cite{Bruenn1985}, in the form of Eq.~(32) of Ref.~\cite{Reddy1998}. For $\bar\nu_ep\to e^+n$ we use $E_{e^+}=\omega-Q$, positron blocking and $\eta_{pn}$. The transport scattering opacity on nucleons is
\begin{align}
\kappa_s=\sigma_0\left(\frac{\omega}{m_e}\right)^2\Big[&\frac{1+5g_A^2}{24}\,\tilde n_n\nonumber\\
&+\frac{(C_V'-1)^2+5g_A^2(C_A'-1)^2}{6}\,\tilde n_p\Big],
\label{eq:kscat}
\end{align}
with $C_V'=\frac12+2\sin^2\theta_W$, $C_A'=\frac12$, $\sin^2\theta_W=0.23$ \cite{Burrows2004}, and the Pauli-blocked density $\tilde n_i=T\,\partial n_i/\partial\nu_i$ \cite{Reddy1998}. In the semitransparent mantle we cap the diffusion coefficients as $D_n\to D_n/(1+\lambda/r)$, where $\lambda=D_4/D_4^{(0)}$ and $D_4^{(0)}=6\times7\pi^4/30$ is the value of $D_4$ for unit mean free path and $\eta=0$. Equation~(\ref{eq:kabs}) is the elastic limit, without the weak-magnetism and recoil factor of Ref.~\cite{Burrows2004}, which lowers the $\bar\nu_e$ absorption opacity by about 15\% at 20 MeV; Ref.~\cite{Pons1999} uses the full rates of Ref.~\cite{Reddy1998}. The cap on $D_n$ is our heuristic, not one of the flux limiters of Ref.~\cite{Pons1999}. We omit neutrino-electron scattering and the pair processes, so for $\nu_{\mu,\tau}$ the opacity is scattering on nucleons alone; with the values of $G_F$ and $|V_{ud}|$ of today, $\sigma_0$ would be 3\% larger for the neutral-current and 2\% smaller for the charged-current processes, far less than the factor-2 opacity variations of our evolutions. At the surface the fluxes are free-streaming, $F_\nu=\alpha n_\nu$ and $H_\nu=\alpha J_\nu$ with $\alpha=0.5$, inside the range 0.4--0.6 of Ref.~\cite{Pons1999} [their Eq.~(41)].

The initial model has $s=1$ and $Y_L=0.36$ in the core and $s=8$ and $Y_L=0.25$ in the shocked mantle, joined at 53\% of the baryon mass by a $\tanh$ profile of width 0.04. The central values, the mantle entropy and the start of the shocked mantle near half the baryon mass follow Ref.~\cite{Pons1999}, whose initial models have $Y_L\sim0.36$ and $s\sim1$ in the center and $s\ge8$ in a mantle that begins at about 0.5 $M_\odot$ of enclosed baryon mass in a star of 1.08 $M_\odot$, with the profiles scaled to the relative mass for heavier stars; the mantle lepton fraction, the $\tanh$ form and its width are our choice. The lepton fraction of the mantle beyond 56\% of the baryon mass is moved to its value without trapped neutrinos ($\mu_\nu=0$), linearly over a further 5\% of the mass. We use 151 zones, uniform in $a$ in the inner 80\% of the mass and refined toward the surface. Each time step solves Eqs.~(\ref{eq:YL}) and (\ref{eq:s}) implicitly at fixed structure, and then relaxes the structure, Eq.~(\ref{eq:struct}), at fixed $s$ and $Y_L$ in each zone. Reference~\cite{Pons1999} iterates these two steps to convergence (3--4 iterations); we make one pass per step and control the error through the step size. The baseline has baryon mass 1.45 $M_\odot$; the eleven variants change the core entropy (0.7, 1.5), the mantle entropy (6, 10), the opacities (factors 0.5, 2), the mass at which the mantle begins ($\pm$0.1 in the mass fraction), and the equation of state (FSU2R, with two further variants of opacity and mantle entropy). The core entropies 0.7 and 1.5 lie within the range that Ref.~\cite{Pons1999} expects, an entropy 50\% larger or smaller than $s\approx1$.

Like those of Ref.~\cite{Pons1999}, our models are spherical and have no accretion, rotation, magnetic fields or convection. Spherical models describe the deleptonization of the PNS well \cite{Fischer2021axion}, but much of the star becomes convectively unstable after a few seconds, and convection would raise the neutrino luminosities and speed up the cooling \cite{Pons1999}. Our evolutions also omit the energy carried away by ALPs; for axions with couplings compatible with current bounds, Fischer \textit{et al.} find that this energy loss lowers the neutrino luminosity significantly during the first 10 s \cite{Fischer2021axion}. To size this effect for our ALPs, we computed the ALP luminosity $L_a(t)$ of each evolution with the baseline rate plus pion conversion at $g_{ap}=2.7\times10^{-10}$ GeV$^{-1}$, letting every ALP escape. At $m_a=10$ MeV the ALPs carry at most \EaEnuTen\% of the energy radiated in neutrinos over 10 s, but near $m_a=\LaLnuPeakMass$ MeV $L_a$ reaches \LaLnuPeak\% of the neutrino luminosity $L_\nu$ and the ALPs carry up to \EaEnuMax\% of the neutrino energy. These are upper limits, because absorption in the PNS (Sec.~\ref{sec:surv}) removes part of the ALPs, but they show that near 100 MeV the ALP energy loss would cool the PNS faster and lower the later production. Near 100 MeV a self-consistent evolution that includes the ALP energy loss is therefore needed to fix the survivor fractions; at 10 MeV the omission is small.

\subsection{The bound population}
\label{sec:formal_bound}

From these profiles and the production rates we determine which ALPs become bound and on which orbits. Reference~\cite{Bose2026ghosts} assigns an ALP produced at radius $r$ with speed $v$ and tangential speed $v_\phi$ to the radius $r'$ at which its radial velocity vanishes in the Newtonian potential of the PNS [their Eqs.~(7) and (8)], and counts the bound density with the velocity measure of their Eq.~(A1), $d^3p_a=2m_a^3v^2(v^2-v_\phi^2)^{-1/2}dv\,dv_\phi$. Integrated over $0<v_\phi<v$ this measure gives $\pi m_a^3v^2dv$, while the isotropic measure, summed over the two signs of the radial velocity and over the azimuth of the tangential velocity,
\begin{equation}
d^3p_a=4\pi m_a^3\frac{v\,v_\phi}{\sqrt{v^2-v_\phi^2}}\,dv\,dv_\phi,
\label{eq:measure}
\end{equation}
gives $4\pi m_a^3v^2dv$. The counting of Ref.~\cite{Bose2026ghosts} therefore holds about a quarter of the bound ALPs and favors nearly radial orbits. It also places each ALP at a single radius, so it cannot follow an orbit through the star. We instead follow each ALP along its orbit, as in calculations of stellar basins, where a bound orbit contributes to the density at each radius in proportion to the time it spends there \cite{VanTilburg2021,Berlin2022helio}.

In the metric (\ref{eq:metric}) of each time slice of our PNS evolution, with $\phi$ from $d\phi/dr=(m+4\pi r^3P)/[r(r-2m)]$, an ALP of conserved energy $E$ and angular momentum $L$ per unit mass obeys
\begin{equation}
e^{2\Lambda}\left(\frac{dr}{d\tau}\right)^2=E^2e^{-2\phi}-1-\frac{L^2}{r^2}\equiv\mathcal{F}(r),
\label{eq:orbit}
\end{equation}
and it is bound when $E<1$. A static observer at $r$ sees the speed $v$ with $\gamma=Ee^{-\phi}$. An ALP produced at $r_0$ with speed $v_0$ at angle $\theta_0$ to the radial direction has $E=\gamma_0e^{\phi(r_0)}$ and $L=r_0\gamma_0v_0\sin\theta_0$. Its turning points $r_p<r_a$ are the zeros of $\mathcal{F}$. The PNS contracts on a time scale of 0.1--1 s, long compared with orbital periods of milliseconds, so we hold $L$ and the radial action
\begin{equation}
J_r=\int_{r_p}^{r_a}e^{\Lambda}\sqrt{\mathcal{F}(r)}\,dr
\label{eq:Jr}
\end{equation}
fixed, since both are adiabatic invariants \cite{Landau1976}. The derivative $\partial J_r/\partial E$ equals the coordinate time of a half orbit, from $r_p$ to $r_a$, which we use in a Newton iteration for $E$ on each new slice. We tested this map against five orbits integrated directly in Newtonian gravity through a contracting star: the final periapses agree to within 3 m.

ALPs are produced on each slice according to Eq.~(\ref{eq:prod}), with the local proper volume $4\pi r^2e^{\Lambda}dr$, the local time $d\tau=e^{\phi}dt$, and an isotropic direction distribution. While the PNS is hot an ALP survives with probability
\begin{equation}
P_{\rm surv}=\exp\left[-\int dt\,\left\langle\Gamma_{\rm abs}\big(1-e^{-\omega/T}\big)e^{\phi}\right\rangle_{\rm orbit}\right],
\label{eq:Psurv}
\end{equation}
where the orbit average weights each radius with the coordinate time spent there and $\omega=\gamma m_a$ is the ALP energy measured by a static observer at that radius. The factor $1-e^{-\omega/T}$ subtracts the stimulated emission, so that the net rate is consistent with Eq.~(\ref{eq:db}) \cite{VanTilburg2021,Lella2024}. After the PNS phase the star is cold, and every orbit with periapsis inside the cold radius $R$ is absorbed (Sec.~\ref{sec:cold}); an ALP survives only if $r_p(10\,{\rm s})>R$. We count ALPs whose final apoapsis is at most 100 km, as Ref.~\cite{Bose2026ghosts} does. We sample $2\times10^4$ ALPs for each of two random seeds, with importance sampling in radius and direction, on 18 time slices between 0.1 and 10 s, at twelve masses between 3 and 400 MeV. For every ALP we evaluate the cold-star cut, the capture cone below and the redshift on a grid of cold radii $R=10.5$--14 km in steps of 0.25 km, so the same sample gives the heating for any final radius.

\subsection{Heating and limits}
\label{sec:formal_limits}

The ALPs that survive heat the old star through their decays. Each decay gives two photons of energy $m_a/2$ in the ALP frame. A photon emitted at radius $r$ heats the star if it reaches the surface, with probability $\Omega(r)/4\pi$, as in Ref.~\cite{Bose2026ghosts} [their Eq.~(10)]. Summed over the ALPs that we follow, the heating luminosity is
\begin{equation}
L_{\rm NS}=\frac{m_a}{\tau_a}\sum_iP_{{\rm surv},i}\left\langle e^{2\phi}\frac{\Omega}{4\pi}\right\rangle_i\exp\left[-\frac{t}{\tau_a}\left\langle\frac{e^{2\phi}}{E_i}\right\rangle_i\right],
\label{eq:LNS}
\end{equation}
where $E_i$ is the conserved energy per unit mass of ALP $i$ and the averages are over coordinate time on its orbit. An ALP decays at the rate $e^{2\phi}/(E_i\tau_a)$ per unit coordinate time, and its photons carry the energy $m_aE_i$ to infinity, so its luminosity at infinity is proportional to $e^{2\phi}\Omega/4\pi$ and independent of $E_i$. Reference~\cite{Bose2026ghosts} [their Eq.~(10)] counts $m_a/2$ per decay, which is half of Eq.~(\ref{eq:LNS}). Unlike the flat-space solid angle of Ref.~\cite{Bose2026ghosts} [their Eq.~(10)], our captured fraction includes light bending: a photon emitted isotropically at $r>R$ reaches the surface if its impact parameter is below $b_c=Re^{-\phi(R)}$ \cite{Pechenick1983}, so
\begin{equation}
\frac{\Omega(r)}{4\pi}=\frac12\left[1-\sqrt{1-\frac{b_c^2e^{2\phi(r)}}{r^2}}\right],
\label{eq:OmegaGR}
\end{equation}
and $\Omega=4\pi$ inside the star. Since $R>3M$, a photon emitted outward escapes and an inward photon reaches the surface if and only if its impact parameter is below $b_c$, so Eq.~(\ref{eq:OmegaGR}) is exact for a static emitter; we neglect the aberration due to the motion of the ALP. Aberration beams the photons forward along the orbit, and for the nearly circular orbits that survive this direction is tangential, so the static cone slightly overestimates the heating. We take $\phi$ from the PNS at 10 s. For $R$ we take the radius of the 1.35 $M_\odot$ cold star built from the same RMF model as the PNS, solved at $T=0.5$ MeV in beta equilibrium without neutrinos and joined to the SLy crust \cite{Douchin2001}: \RcoldGM\ km for GM3 and \RcoldFS\ km for FSU2R. We convert to the surface luminosity with $L_{\rm surf}=L_\infty/(1-2M/R)$, as in Table I of Ref.~\cite{Bose2026ghosts}. Without absorption $P_{\rm surv}=1$ and the cold-star condition is dropped. We call $R_L$ the ratio of Eq.~(\ref{eq:LNS}) with absorption to that without.

With $x=t/\tau_a$ and $\mathcal{N}$ the weighted number of bound ALPs, Eq.~(\ref{eq:LNS}) is close to $L_{\rm NS}=m_a\mathcal{N}x\,e^{-x}/t$; the slowing of the decay, $\langle e^{2\phi}/E_i\rangle_i\simeq$ \DecSlow\ for the survivors, shifts $x$ slightly, and we evaluate the full sum. We attribute the whole observed thermal luminosity $L_{\rm obs}$ to the ALP decays, as in studies of the heating of old NSs by dark matter, where the heating is balanced by the blackbody emission $4\pi R^2\sigma_{\rm SB}T^4$ \cite{Baryakhtar2017}. Other heating mechanisms would leave less room for the ALPs; a model with rotochemical heating with a large Cooper-pairing gap and vortex creep predicts a temperature of PSR J2144$-$3933 close to its upper limit \cite{Rodriguez2026heating}. Our choice is therefore conservative. The limit $L_{\rm NS}\le L_{\rm obs}$ at age $t$ then excludes $x_1<x<x_2$, where $x_{1,2}=-W_{0,-1}(-1/A)$ are the two real branches of the Lambert function and
\begin{equation}
A=\frac{m_a\,\mathcal{N}}{t\,L_{\rm obs}}.
\label{eq:A}
\end{equation}
The edges in coupling are $g_{a\gamma\gamma}=[64\pi x_{1,2}/(m_a^3t)]^{1/2}$, and a window exists only if $A>e$. We call $A/e=\max_g L_{\rm NS}/L_{\rm obs}$ the margin of the window.

For PSR J2144$-$3933 we take, as Ref.~\cite{Bose2026ghosts} does in their Eq.~(11), $L_{\rm obs}=1.4\times10^{27}$ erg s$^{-1}$ for $T=3.4\times10^4$ K, which they derive from the far-ultraviolet flux limit of Ref.~\cite{Guillot2019} for a distance of 192 pc and $E(B-V)=0.06$; this is close to $4\pi R^2\sigma_{\rm SB}T^4$ for $R=11.9$ km, so it is a luminosity at the surface, which we compare with $L_{\rm surf}$. Table I of Ref.~\cite{Bose2026ghosts} lists instead $L_{\rm surf}<1.8\times10^{27}$ erg s$^{-1}$ for this star. We use the value of their Eq.~(11), because it follows from the temperature and radius of their Eqs.~(11) and (12) and because their text derives from it the J2144$-$3933 limits shown in their Fig.~3, although the caption of that figure refers to their Table I; the larger value would lower the J2144$-$3933 margins by the factor 1.8/1.4, which the conservative variant below covers. We take the characteristic age, $t=333$ Myr \cite{Guillot2019}. Reference~\cite{Guillot2019} also gives a conservative bound, $T<4.2\times10^4$ K, for the largest distance and extinction and a radius of 10 km; the corresponding $4\pi R^2\sigma_{\rm SB}T^4$ is larger than that for $T=3.4\times10^4$ K and $R=11.9$ km by a factor \ConsFac, and in a conservative variant (Table~\ref{tab:stars}) we multiply $L_{\rm obs}$ by this factor. A reanalysis with the recalibrated Hubble Space Telescope photometry gives a redshifted surface temperature below $3\times10^4$ K \cite{Rodriguez2026heating} [their Table~1]; for radii of 10--13 km it corresponds to a luminosity between our two values.

Reference~\cite{Bose2026ghosts} fixes $g_{ap}=2.7\times10^{-10}$ GeV$^{-1}$, consistent with the SN 1987A cooling bound of Lella \textit{et al.}, who computed it with bremsstrahlung and pion conversion; for $m_a\lesssim10$ MeV that bound excludes $g_{ap}\gtrsim6\times10^{-10}$ in their dimensionless normalization, $C_pm_N/f_a$, i.e. $g_{ap}\gtrsim3.2\times10^{-10}$ GeV$^{-1}$ in ours \cite{Lella2024}, so it is set by ALPs light compared with the temperature. The value of $g_{ap}$ is therefore tied to a production rate. To compare rates at the same distance from the bound, we run the two alternative rates with $g_{ap}^2$ rescaled so that the ALP energy that the PNS radiates over 0.1--10 s at $m_a=3$ MeV equals that of the baseline. This multiplies the $pp$-only rate by \ppScaleGM\ (GM3) and \ppScaleFS\ (FSU2R), and the OPE rate by \opeScaleGM\ (GM3) and \opeScaleFS\ (FSU2R). The same argument applies to the effective mass: a model that lowers the rate also weakens the cooling bound, and the two changes largely compensate in the heating. For the limits we use, at each point of the $(m_a,g_{a\gamma\gamma})$ plane, the smallest heating among our PNS evolutions; we interpolate the heating in $m_a$ with a monotone cubic in $\ln m_a$ between the twelve masses.

%=====================================================================
\section{Absorption in the cold star}
\label{sec:cold}
%=====================================================================

We first ask how fast the cold star absorbs a bound ALP whose orbit enters it. In the cold star the rate does not depend on temperature, because $\omega\gg T$. An ALP absorbed with energy $\omega$ moves two nucleons from the Fermi sea into empty states up to $\omega$ above the Fermi surface, so Pauli blocking limits the final states but does not suppress the rate exponentially. Nucleon pairing changes the spectrum only within a gap of order an MeV or less of the Fermi surface; for neutron $^1S_0$ pairing the calculated critical temperature is at most (0.5--0.7)$\times10^{10}$ K \cite{Page2004}. Gaps of this size play no role for $\omega\ge10$ MeV.

We compute $\Gamma_{\rm abs}(r)$ from Eq.~(\ref{eq:gammaC}) at nine radii in the core ($n_B>0.08$ fm$^{-3}$) of 1.35 $M_\odot$ stars for the 19 RMF sets, obtained from the Tolman-Oppenheimer-Volkoff equations with beta-equilibrated matter at $T=0$ joined to the SLy crust \cite{Douchin2001}, for all three rates. We take a radial orbit with apoapsis $r_a=50$ km, so that $E=(1-2M/r_a)^{1/2}$, the local Lorentz factor is $\gamma(r)=Ee^{-\phi(r)}$, $\omega=\gamma m_a$, and the period is the Kepler period of a radial orbit with semimajor axis $r_a/2$, $P_{\rm orb}=2\pi(r_a/2)^{3/2}(GM)^{-1/2}$. The optical depth per crossing is
\begin{equation}
\tau_1=2\int_0^{R}\frac{\Gamma_{\rm abs}(r)\,dr}{\zeta(r)\sqrt{1-2m(r)/r}},
\label{eq:tau1}
\end{equation}
where $e^{\Lambda}dr/\zeta=dr/[\zeta(1-2m/r)^{1/2}]$ is the time in which a static observer at $r$ sees the ALP cross $dr$. We integrate over the core only and omit the crust, which underestimates $\tau_1$. Since $\tau_1\ll1$ in all cases, the ALP crosses the star many times, once per period, and it is absorbed after $t_{\rm abs}=P_{\rm orb}/\tau_1$. Because $P_{\rm orb}\propto r_a^{3/2}$, an orbit with apoapsis 100 km, the largest that we count, has a period, and so approximately an absorption time, \TabsApoFac\ times longer.

\begin{figure*}[t]
\includegraphics[width=\textwidth]{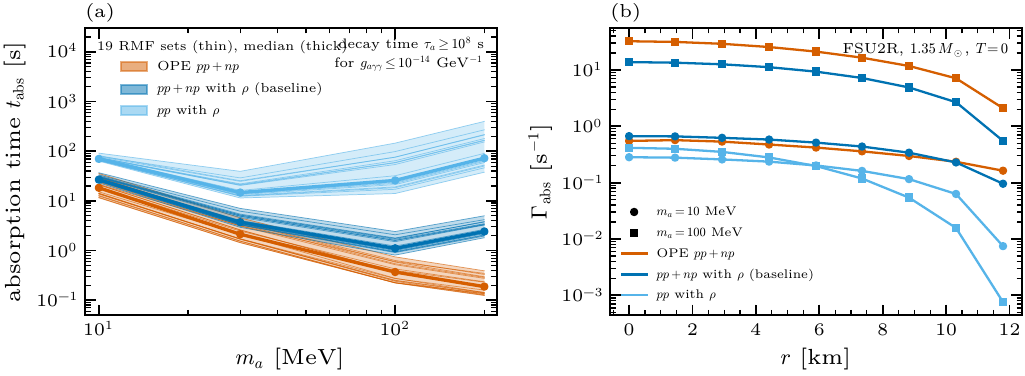}
\caption{Absorption in a cold 1.35 $M_\odot$ NS. (a) Absorption time $t_{\rm abs}=P_{\rm orb}/\tau_1$ of a bound ALP on a radial orbit with apoapsis 50 km, at $m_a=10$, 30, 100 and 200 MeV, for the 19 RMF sets (thin lines; band: their range; thick line with symbols: median) and three rates: OPE $pp+np$ (vermilion), $pp+np$ with $\rho$ exchange (blue, our baseline), and $pp$ with $\rho$ exchange (light blue, the process of Ref.~\cite{Bose2026ghosts}). The label gives a lower bound on the decay time $\tau_a$ for $g_{a\gamma\gamma}\le10^{-14}$ GeV$^{-1}$ at these masses. (b) Absorption rate $\Gamma_{\rm abs}$, measured by a static observer, at the nine radii of the core ($n_B>0.08$ fm$^{-3}$) of the FSU2R star for the same orbit, at $m_a=10$ (circles) and 100 MeV (squares), for the same three rates.}
\label{fig:abs}
\end{figure*}

In Fig.~\ref{fig:abs} we show the absorption time and the absorption rate in the cold star. We see that a star-crossing ALP is absorbed in \TabsOPE\ s with OPE, in \TabsBase\ s with our baseline, and in \TabsPP\ s with the $pp$ channel alone, at $m_a=10$--200 MeV. With OPE the rate grows with $\omega$, so the median $t_{\rm abs}$ falls over the whole mass range. With $\rho$ exchange the median $t_{\rm abs}$ has a minimum, at \TabsMinMassBase\ MeV for the baseline and at \TabsMinMassPP\ MeV for the $pp$ channel, because the $\rho$ term of Eq.~(\ref{eq:rho}) lowers the rate below OPE by a factor that grows with $\omega$. The 19 equations of state change $t_{\rm abs}$ at fixed mass by a factor \EosFacBase\ for the baseline and \EosFacOPE\ for OPE, and by up to \EosFacPP\ for the $pp$ channel; the largest stars absorb fastest (correlation coefficient \RtabsCorr\ between $R$ and $t_{\rm abs}$ for the baseline), because their proton fraction is higher. Across the 19 sets $t_{\rm abs}$ follows the mean proton fraction of the core more closely than the radius. Adding the $np$ channel shortens $t_{\rm abs}$ by a factor \npColdTen\ at 10 MeV and by \npColdHi\ at 100--200 MeV (medians \npColdHiMed), because the $pp$ rate falls steeply with $\omega$ in the outer core, where the protons are few, while the charged-pion term of Eq.~(\ref{eq:Mnp}) does not. For the couplings of the surviving windows below, $g_{a\gamma\gamma}\le$ \UnionGmax\ GeV$^{-1}$, the decay time is at least \TauMin\ s, so absorption is faster than decay by more than \AbsDecayOrders\ orders of magnitude for every rate and equation of state, and for apoapses up to 100 km.

$\Delta$ isobars change the cold star but not this conclusion. We have repeated the calculation with the $\Delta$ quartet of Eq.~(\ref{eq:LDelta}) for GM3 and FSU2R with $x_\sigma=1.0$, 1.1 and 1.18. In GM3 the $\Delta^-$ appears in a 1.35 $M_\odot$ star only for $x_\sigma=1.18$, with \DcoldGMfrac\% of the baryons at the center, and $t_{\rm abs}$ changes by at most \DcoldGMchange\%. In FSU2R with $x_\sigma=1.0$, \DcoldFSfrac\% of the central baryons are $\Delta^-$, the radius falls from \RcoldFS\ to \RcoldDtwo\ km, and the baseline $t_{\rm abs}$ shortens by a factor \DcoldFSfac, because the $\Delta^-$ raise the proton fraction of the nucleons, as Refs.~\cite{Sedrakian2022Delta,Parmar2025} also find. This star is smaller and absorbs faster, so it is the proton fraction, not the radius, that sets $t_{\rm abs}$. These two $\Delta$-admixed models have maximum masses of \MmaxGMD\ and \MmaxFSD\ $M_\odot$, so we use them as tests of the sensitivity rather than as realistic models. The GM3 model with $x_\sigma=1.0$ reaches \MmaxGMDone\ $M_\odot$ but has no $\Delta$ in a 1.35 $M_\odot$ star.

The conclusion of this section is that every bound ALP whose orbit enters the cold star is absorbed long before it decays. Its energy goes into the young star, which is still hot, and it does not heat the old one. This removes in particular the ALPs whose orbits lie entirely inside the star; in Ref.~\cite{Bose2026ghosts} their decays give a heating $O(1)$ times greater than that from the ALPs outside the star.

%=====================================================================
\section{Which ALPs survive}
\label{sec:surv}
%=====================================================================

The question is then how many ALPs are on orbits whose periapsis stays outside the final star. An ALP produced at radius $r$ has its periapsis at or inside $r$. A survivor must therefore be produced outside the final radius, with a mostly tangential velocity, and it must avoid the hot PNS matter while the PNS is still extended. All three conditions depend on how the PNS evolves. We compute the evolution ourselves, with the finite-temperature RMF equation of state of Eqs.~(\ref{eq:LRMF})--(\ref{eq:beta}) and the transport of Eqs.~(\ref{eq:struct})--(\ref{eq:kscat}), for GM3 \cite{Glendenning1991} and FSU2R \cite{Tolos2017}.

\begin{figure*}[t]
\includegraphics[width=\textwidth]{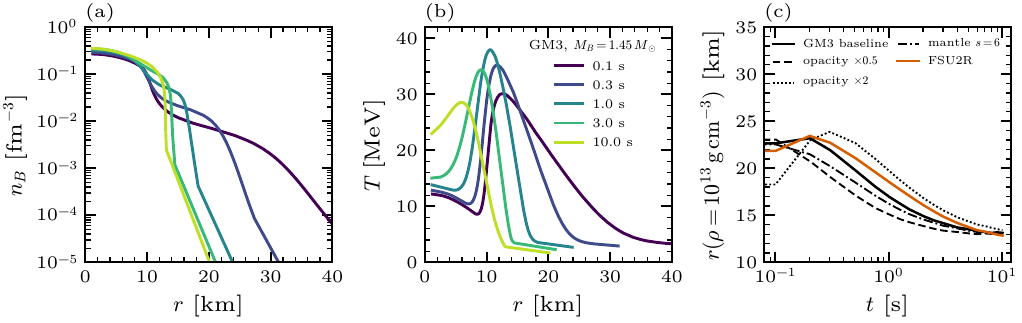}
\caption{Our PNS evolution. (a) Baryon density and (b) temperature profiles at 0.1, 0.3, 1, 3 and 10 s for GM3 with baryon mass 1.45 $M_\odot$, our baseline evolution. (c) Radius at which the density falls to $10^{13}$ g cm$^{-3}$, as a function of time, for five of our evolutions: the GM3 baseline, its variants with the opacities multiplied by 0.5 and 2 and with mantle entropy $s=6$, and the FSU2R baseline.}
\label{fig:pns}
\end{figure*}

In Fig.~\ref{fig:pns} we show the evolution. We see that in the GM3 baseline the radius at $10^{13}$ g cm$^{-3}$ shrinks from about \PnsRb\ km at 0.2 s to about \PnsRc\ km at 1 s, and then more slowly; at 10 s it is \PnsRten\ km in the twelve evolutions. As in the evolutions of Pons \textit{et al.}, the outer mantle contracts during the first few tenths of a second because it loses pressure by neutrino emission \cite{Pons1999}. The compression heats the mantle \cite{Pons1999}, so in our evolutions the temperature peaks off center. The peak first rises to \PnsTpeak\ MeV in the twelve evolutions and then moves inward, while neutrino diffusion heats the center (Joule heating) \cite{Pons1999}. At 10 s the gravitational mass is \PnsMgten\ $M_\odot$, close to the 1.35 $M_\odot$ at 10 s of the SFHo-18.8 simulation used in Ref.~\cite{Bose2026ghosts}, in which the radius that encloses this mass is 11.9 km [their Eq.~(12)]. The twelve evolutions differ in the ways listed at the end of Sec.~\ref{sec:formal_pns}.

For all twelve evolutions we follow each bound ALP with Eqs.~(\ref{eq:orbit})--(\ref{eq:Psurv}) and count the heating with Eqs.~(\ref{eq:LNS}) and (\ref{eq:OmegaGR}) for the baseline rate. For the two alternative rates we use the GM3 and FSU2R baseline evolutions.

\begin{figure*}[t]
\includegraphics[width=\textwidth]{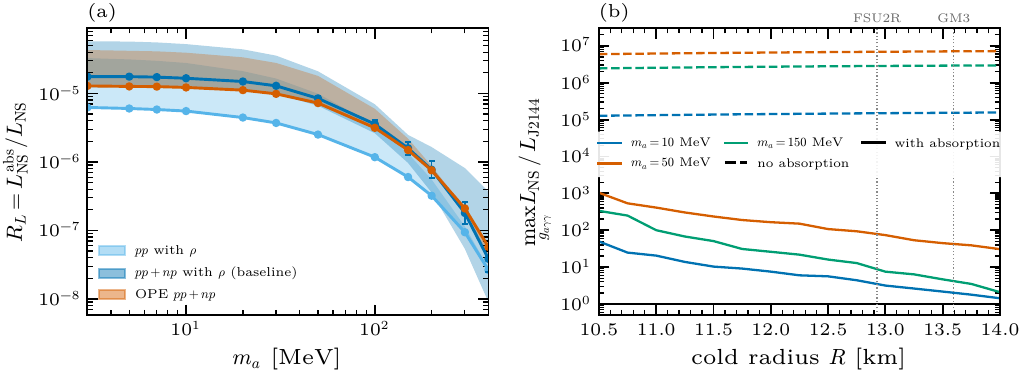}
\caption{Survival of the heating. (a) Fraction $R_L$ of the heating that survives absorption, as a function of $m_a$, for our baseline rate (blue, twelve PNS evolutions) and for the $pp$-only (light blue) and OPE (vermilion) rates at the same light-ALP supernova energy loss (GM3 and FSU2R baseline evolutions). Bands span the evolutions, lines with symbols are the GM3 baseline, and the error bars are its Monte Carlo errors from two seeds. Each evolution uses the radius of the cold star of its own equation of state. (b) Largest heating over $g_{a\gamma\gamma}$, in units of the luminosity $L_{\rm obs}=1.4\times10^{27}$ erg s$^{-1}$ of PSR J2144$-$3933 at $t=333$ Myr ($L_{\rm J2144}$ on the axis), without absorption (dashed) and with it (solid), as a function of the cold radius $R$ common to all evolutions, at $m_a=10$, 50 and 150 MeV, for the smallest heating over the twelve evolutions. A window exists only where a curve lies above the horizontal line at unity; the vertical lines mark the cold radii of GM3 and FSU2R.}
\label{fig:surv}
\end{figure*}

In Fig.~\ref{fig:surv}(a) we show the ratio $R_L$ of the heating with absorption to that without. We see that it is \RLrange\ and that it falls with the ALP mass. It is small for three reasons. First, the production rate scales roughly as $n_B^2$, so most ALPs are born deep in the core and their periapsis is small. Second, the contraction pulls periapses inward. Third, orbits whose periapsis lies inside the hot mantle are absorbed during the first seconds. The surviving fraction is set mainly by the orbits and not by the size of the rate, so the three rates give $R_L$ within a factor \RLrateFac\ of each other. Among the twelve evolutions $R_L$ varies by a factor \RLspread\ below 200 MeV. Part of this spread is the cold radius: the FSU2R evolutions end in a star of \RcoldFS\ km and the GM3 evolutions in one of \RcoldGM\ km.

The surviving heating is carried by a few ALPs on wide, nearly circular orbits, so the Monte Carlo sum has a small effective sample size. From the difference between the two seeds we estimate an error of up to \MCerr\% (median \MCerrMed\%) in the surviving heating of one evolution at $m_a\le200$ MeV. The envelope over twelve evolutions takes the smallest heating at each point and is therefore biased low by downward fluctuations, which is conservative.

In Fig.~\ref{fig:surv}(b) we show the largest J2144$-$3933 heating when all twelve evolutions end in a star of the same radius $R$. We see that the final radius is the main structural uncertainty. Over all masses, the largest J2144$-$3933 heating is \RsensHi\ times the limit for $R=\RsensRlo$ km, \RsensMid\ for 12.5 km and \RsensLo\ for $R=\RsensRhi$ km, and the window spans \RsensmHi\ MeV and \RsensmLo\ MeV at the two ends. A smaller star lets more orbits clear it. This range brackets the radii of our 19 cold RMF stars, \RcoldRange\ km, between which the margin changes by a factor \RsensRatioSet; the cold GM3 star, at \RcoldGM\ km, is slightly larger. The radius of PSR J2144$-$3933 is not measured, and we use the radius of the cold star built from the same equation of state as the PNS.

Pion conversion, Eq.~(\ref{eq:gpion}), changes the survivors above $m_\pi$. We have repeated the survival calculation for all twelve evolutions at $m_a=150$--400 MeV with bremsstrahlung plus pion conversion, including the $\Delta$ resonance. The heating without absorption grows by a factor \PiProdFac, and the surviving heating by \PiSurvFac, although the hot absorption is now much stronger. Pion conversion is a one-body process in the nucleon density, whereas bremsstrahlung goes as $n_B^2$; we attribute the larger survival to the larger share of ALPs born in the low-density mantle, where the survivors come from. At 300--400 MeV the $\pi^0$ channels, mainly $\pi^0N\to\Delta\to Na$, give about half of the pion-conversion production (Sec.~\ref{sec:formal_brems}).

$\Delta$ isobars act mainly through the radius of the final star. In hot PNS matter the $\Delta$s appear gradually. In our GM3 evolution with $x_\sigma=1.18$ (D1) they make up at most \DoneFirst\% of the baryons during the first second, when the survivors are born, and \DoneCenter\% at the center at 10 s. In the FSU2R evolution with $x_\sigma=1.0$ (D2) they reach \DtwoCenter\% at the center at 10 s. With the radii of the corresponding $\Delta$-admixed cold stars, \RcoldDone\ and \RcoldDtwo\ km, $R_L$ changes by a factor \DoneRL\ for D1 and \DtwoRL\ for D2 relative to their nucleonic partners. The largest J2144$-$3933 heating is \BOneMargin\ times the limit for GM3 without $\Delta$s and \DoneMargin\ with them, and changes from \FOneMargin\ to \DtwoMargin\ for FSU2R.

%=====================================================================
\section{Heating limits}
\label{sec:limits}
%=====================================================================

\begin{figure}[t]
\includegraphics[width=\columnwidth]{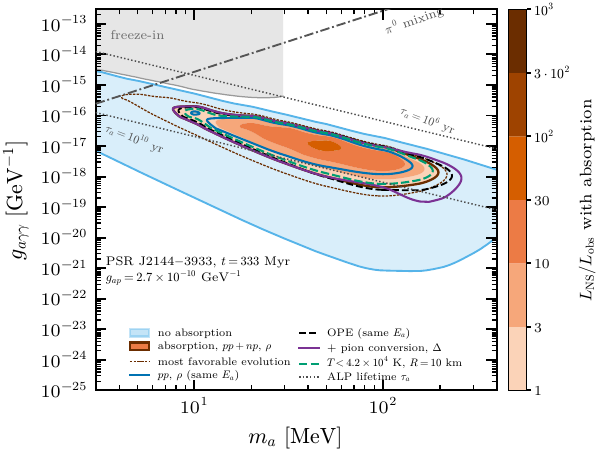}
\caption{Heating limit from PSR J2144$-$3933 ($L_{\rm obs}=1.4\times10^{27}$ erg s$^{-1}$ \cite{Bose2026ghosts}, age 333 Myr \cite{Guillot2019}) for $g_{ap}=2.7\times10^{-10}$ GeV$^{-1}$. Filled contours: $L_{\rm NS}/L_{\rm obs}$ with absorption for our baseline rate, smallest over twelve PNS evolutions, each with the cold radius of its own equation of state; thick brown line: $L_{\rm NS}=L_{\rm obs}$. Thin brown dotted line: the same edge for the most favorable evolution. Light blue: the region excluded without absorption, for the smallest heating. Lines: the edge with absorption for the $pp$-only (blue) and OPE (black dashed) rates at the same light-ALP supernova energy loss, smallest over the GM3 and FSU2R baseline evolutions, and for the baseline with the conservative temperature bound (green dashed). Purple: the edge for the baseline plus pion conversion and the $\Delta$ resonance, Eq.~(\ref{eq:gpion}). Gray: irreducible freeze-in background \cite{Langhoff2022}, from the AxionLimits data \cite{OHare2020}, which extend to 30 MeV. Dash-dotted: the photon coupling induced by ALP-pion mixing, Eq.~(\ref{eq:gind}). Gray dotted lines: ALP lifetimes $\tau_a=10^6$ and $10^{10}$ yr, as labeled. The heating is computed at twelve masses and interpolated in $m_a$.}
\label{fig:J2144}
\end{figure}

We now turn the surviving heating into limits on $g_{a\gamma\gamma}$, first for PSR J2144$-$3933, the coldest NS, and then for the other stars of Ref.~\cite{Bose2026ghosts}. In Fig.~\ref{fig:J2144} we show what absorption does to the J2144$-$3933 limit. We see that the largest heating drops from \JnoAbsMax\ times the observed limit to \JabsMax\ times it with our baseline rate. A window survives for every evolution. For the smallest heating among the evolutions it lies at \JmRange\ MeV and $g_{a\gamma\gamma}=$ \JgRange\ GeV$^{-1}$, with a margin of \Jmargin. At the computed masses inside the window its lower edge lies \JedgeRise\ times above the edge without absorption, at $g_{a\gamma\gamma}\simeq$ \JedgeTen\ GeV$^{-1}$ at 10 MeV and \JedgeHundred\ GeV$^{-1}$ at 100 MeV, and its ceiling, set by decay within the age of the star, falls by a factor \JceilFall. The most favorable evolution extends the window to \JmRangeFav\ MeV with a margin of \JmarginFav. The conservative temperature bound narrows the window to \JmRangeCons\ MeV, with a margin of \JmarginCons; this variant also covers the larger luminosity of Table I of Ref.~\cite{Bose2026ghosts} and the recalibrated temperature limit of Ref.~\cite{Rodriguez2026heating} (Sec.~\ref{sec:formal_limits}).

At the same light-ALP energy loss the $pp$-only rate gives a window at \ppmRange\ MeV with a margin of \ppMargin, and OPE gives one at \opemRange\ MeV with a margin of \opeMargin, against \baseTwoMargin\ for the baseline over the same two evolutions, the GM3 and FSU2R baselines. Once $g_{ap}$ is tied to the same light-ALP energy loss, the three rates give margins within a factor \rateSpread\ of each other, while their cold absorption times differ by more than an order of magnitude. The $pp$-only rate gives the smallest margin: its larger $g_{ap}^2$ raises the absorption in the hot PNS as much as the production, and without the $np$ channel the production falls faster with $m_a$.

With pion conversion and the $\Delta$ resonance the J2144$-$3933 window extends to \JmRangePi\ MeV with a margin of \JmarginPi\ (\JmUpperFavPi\ MeV for the most favorable evolution), and the upper edges of the other windows move to \PiUpperOthers\ MeV; the windows below $m_\pi$ are unchanged, because the new channels open only above $m_\pi$.

\begin{table*}[t]
\caption{Windows excluded with absorption for the baseline rate. Mass range, coupling range and margin are for the smallest heating over the twelve PNS evolutions; the last column gives the margin for the most favorable evolution. Luminosities and ages as in Table I of Ref.~\cite{Bose2026ghosts}, except for J2144$-$3933 (their Eq.~(11)) and Cas A (the neutrino luminosity of their Sec.~III). A star marks the alternative inputs discussed in the text: the spin-down age of B0950+08 and the conservative temperature bound of J2144$-$3933.}
\label{tab:stars}
\begin{ruledtabular}
\begin{tabular}{lcccc}
NS & $m_a$ [MeV] & $g_{a\gamma\gamma}$ [GeV$^{-1}$] & margin & max.\\
\hline
% generated by make_numbers_v121.py from v115/scan_v121.json; envelope min over evolutions B1,B2,B3,B4,B5,B6,B7,B8,B9,F1,F2,F3
J2144$-$3933 & 7.7--196 & $4.6\times10^{-19}$--$2.0\times10^{-16}$ & 43 & 180\\
B0950+08 & 5.8--213 & $4.1\times10^{-18}$--$4.1\times10^{-15}$ & 88 & 380\\
J0108$-$1431 & 17--133 & $1.9\times10^{-18}$--$7.9\times10^{-17}$ & 4.8 & 21\\
B1727$-$47 & 15--147 & $8.9\times10^{-17}$--$5.7\times10^{-15}$ & 7.7 & 33\\
J0633+1746 & -- & -- & $<1$ & 1.9\\
J1836+5925 & 16--135 & $1.9\times10^{-17}$--$8.3\times10^{-16}$ & 5.3 & 23\\
Cas A & -- & -- & $<1$ & $<1$\\
B0950+08$^*$ & 14--155 & $4.3\times10^{-18}$--$3.2\times10^{-16}$ & 9.5 & 41\\
J2144$-$3933$^*$ & 8.9--183 & $5.8\times10^{-19}$--$1.6\times10^{-16}$ & 26 & 110\\

\end{tabular}
\end{ruledtabular}
\end{table*}

In Table~\ref{tab:stars} we give the windows of all seven NSs of Ref.~\cite{Bose2026ghosts}. For PSR B0950+08 the surface luminosity $1.2\times10^{29}$ erg s$^{-1}$ of Ref.~\cite{Bose2026ghosts} is consistent with the temperature of Ref.~\cite{Abramkin2022}, and the kinematic age of 1.9 Myr is close to the most probable kinematic age, about 2 Myr, that Ref.~\cite{Abramkin2022} quotes from Igoshev's analysis; the spin-down age is 17.5 Myr \cite{Pavlov2017}. For Cas A, Ref.~\cite{Bose2026ghosts} uses the minimal-cooling neutrino luminosity $3.7\times10^{35}$ erg s$^{-1}$, $L\propto T^8$ \cite{Page2004}, at an internal temperature of $2.8\times10^8$ K. The relations of Ref.~\cite{Heinke2010} for modified Urca emission and an iron envelope give $(8$--$14)\times10^{35}$ erg s$^{-1}$ for the observed surface temperature, and pair breaking can raise it by a factor 10 or more if the core nucleons are superfluid \cite{Heinke2010}. A larger luminosity would only lower the Cas A margin further. The quantity that sets a window is the product of luminosity and age, because at fixed mass the largest heating over $g_{a\gamma\gamma}$ falls as $1/t$ (Sec.~\ref{sec:formal_limits}). PSR B0950+08, whose product is half that of J2144$-$3933, keeps the widest window in mass and the largest margin; its margin falls by a factor \BnineSpinFac\ with the spin-down age. PSR J0108$-$1431, PSR B1727$-$47 and PSR J1836+5925 keep windows, PSR J0633+1746 (Geminga) keeps one only for the most favorable evolutions, and the NS in Cas A, which in Ref.~\cite{Bose2026ghosts} covers, with PSR B0950+08, most of the couplings above the J2144$-$3933 window, keeps none. Together the stars exclude $g_{a\gamma\gamma}\simeq$ \UnionG\ GeV$^{-1}$ at $m_a\simeq$ \UnionM\ MeV for the smallest heating among our evolutions.

%=====================================================================
\section{Discussion}
\label{sec:disc}
%=====================================================================

Three issues decide whether the windows of Sec.~\ref{sec:limits} are new limits.

The first is the existing limits on the same ALP. The irreducible freeze-in background \cite{Langhoff2022} lies above the J2144$-$3933 window where the data extend. Gamma rays from the decay of supernova ALPs constrain the same product of couplings, $g_{ap}g_{a\gamma\gamma}$ \cite{Bose2026ghosts}. Lella \textit{et al.} set limits from SN 1987A, the diffuse background and the ALPs trapped around the NS in Cas A for a proton-coupled ALP at $C_pm_N/f_a=10^{-10}$, i.e. $g_{ap}=5.3\times10^{-11}$ GeV$^{-1}$ \cite{Lella2023}, and Benabou \textit{et al.} set Fermi-LAT limits from time-delayed photons for heavy axions with QCD-like couplings \cite{Benabou2025}. Reference~\cite{Bose2026ghosts} recast the latter to $g_{ap}=2.7\times10^{-10}$ GeV$^{-1}$ and found that its heating limits are often stronger. With absorption the lower edges of the heating windows are two to three orders of magnitude higher, and these gamma-ray limits may cover the surviving windows. The limits of Ref.~\cite{Lella2023} have not been recast to this coupling, and a recast of both sets of gamma-ray limits with the production rate used here is needed to decide. The trapped-ALP limit of Ref.~\cite{Lella2023} for Cas A is itself affected by absorption, since the trapped ALPs cross the NS.

The second is the photon coupling. The windows lie far below the coupling of Eq.~(\ref{eq:gind}). A proton-coupled ALP in the J2144$-$3933 window needs a direct photon coupling that cancels the induced one to one part in \tuneTen\ at 10 MeV and one part in \tuneHundred\ at 100 MeV. This tuning applies equally to the windows of Ref.~\cite{Bose2026ghosts}, whose $g_{a\gamma\gamma}$ is a tuned combination of a direct coupling and the induced coupling [N.~Raj, private communication]. Without it the ALP is excluded outright: with the induced photon coupling of an ALP coupled to light quarks, the diffuse supernova ALP background excludes $g_{ap}\gtrsim8.5\times10^{-11}$ in their normalization $C_pm_N/f_a$, i.e. $g_{ap}\gtrsim4.5\times10^{-11}$ GeV$^{-1}$ in ours, for $10\lesssim m_a\lesssim100$ MeV \cite{Lella2024QCD}, below the value $2.7\times10^{-10}$ GeV$^{-1}$ that we use.

The third is the microphysics and the PNS. Our bremsstrahlung rates are based on one-pion and $\rho$ exchange. Rates built on measured nucleon-nucleon phase shifts are about four times below OPE for soft radiation in neutron matter \cite{Hanhart2001}, and OPE with $\rho$ exchange and multiple-scattering corrections, which our rates do not include, is about 20\% below a vacuum $T$-matrix rate in the inner supernova core \cite{Carenza2019}; we know of no controlled rate for ALP energies of hundreds of MeV. Tying $g_{ap}$ to the supernova energy loss removes most of this uncertainty from the heating, but not from the shape of the survivors' birth region. The nucleon-velocity term of Ref.~\cite{Bai2025}, which we include only in the cold OPE rate, would raise the hot absorption of slow ALPs and lower $R_L$. The one-body channels $aN\to\pi N$ and $aN\to\Delta$ raise the absorption above $m_\pi$ by large factors, but they raise the production of survivors more, so including them widens the windows; we attribute this to the larger share of ALPs born in the low-density mantle (Sec.~\ref{sec:surv}). With vacuum pions our $\pi^-p$ production is likely a lower estimate, because in-medium interactions raise the $\pi^-$ abundance \cite{Fore2020}, and we omit the terms of next order in $1/m$, which lower the pion absorption rate by \PiHoFac\% at 400 MeV; Ho \textit{et al.} find that the large-$m_N$ expansion overestimates the pion-conversion emission \cite{Ho2023}.

Our PNS evolutions start at 0.1 s with a fixed baryon mass, without accretion, convection or muons. In supernova models substantial accretion occurs between 0.1 and 0.5 s \cite{Pons1999}, and Ref.~\cite{Bose2026ghosts} treats the core as accreting for the first second; a growing mass pulls the orbits inward and would lower $R_L$. Pons \textit{et al.} argue that the collapse of an initially extended high-entropy mantle, as in our evolutions, largely simulates the effects of accretion on the PNS \cite{Pons1999}. A calculation on full supernova profiles would settle this. The radius of the final star, which decides which orbits clear it, matters almost as much as the rate: between $R=\RsensRlo$ and \RsensRhi\ km the margin changes by a factor \RsensRatio, and by \RsensRatioSet\ across the radii of our 19 cold stars.

Our treatment of $\Delta$-admixed matter is also incomplete. We include $aN\to\Delta$ only for vacuum nucleons and $\Delta$s in nucleonic matter, and we omit bremsstrahlung on $\Delta$s and the $\Delta$ contributions to the neutrino opacities, and our two $\Delta$-admixed PNS models do not reach $2\,M_\odot$; models with other couplings, such as those of Refs.~\cite{Thapa2021,Parmar2025,Yogeesh2026}, support massive stars with $\Delta$s. None of these issues changes the absorption in the cold star, which removes every star-crossing orbit regardless of the rate, the composition and the PNS history.

The same physics applies to any particle that is bound to a NS and couples to nucleons. For dark photons and for other ALP couplings the absorption channel differs, but the number of crossings is the same. Stable particles produced in pairs, such as the leptophilic freeze-in dark matter of Ref.~\cite{Pospelov2026}, are not removed this way: their rare scatterings at most thermalize and cool the bound population, which is depleted only by annihilation and a small evaporation \cite{Pospelov2026}.

%=====================================================================
\section{Conclusions}
%=====================================================================

We have shown that the bound supernova ALPs of Ref.~\cite{Bose2026ghosts} are absorbed by the star they orbit. An ALP on an orbit that enters the star and reaches 50 km crosses it about \Ncross\ times over the 333 Myr age of PSR J2144$-$3933, and inverse nucleon bremsstrahlung in the cold star absorbs it within \TabsRange. Only ALPs born outside the final radius on nearly circular orbits survive, and they carry \RLrange\ of the heating. With our own PNS evolutions and bremsstrahlung in $pp$ and $np$ collisions, PSR J2144$-$3933 and four other NSs still exclude windows in $g_{a\gamma\gamma}$, and PSR J0633+1746 only for the most favorable evolutions, but their lower edges rise by two to three orders of magnitude. Tied to the same light-ALP supernova energy loss, the three bremsstrahlung rates give J2144$-$3933 margins within a factor \rateSpread, and the radius of the final star changes the margin by a factor \RsensRatioSet\ across our \NsetsPass\ cold stars. Pion conversion and the $\Delta$ resonance raise both the absorption and the production above $m_\pi$, and with them the J2144$-$3933 window extends to \JmRangePi\ MeV. The windows lie far below the photon coupling induced by ALP-pion mixing, which without a cancelling direct coupling would exclude the proton coupling that we use \cite{Lella2024QCD}, and they may be covered by supernova gamma-ray limits. Since the margin grows as the product of luminosity and age falls, deeper temperature limits on old NSs widen the windows. The James Webb Space Telescope, the Extremely Large Telescope and the Thirty Meter Telescope could bound the effective temperature of PSR J2144$-$3933 down to 20000, 15000 and 9000 K, respectively, against the Hubble Space Telescope limit of about 33000 K on the same redshifted temperature for their choice of mass and radius \cite{Raj2024JWST}; with recalibrated photometry the limit is below $3\times10^4$ K \cite{Rodriguez2026heating}.

We see four next steps.
(1) Recast the supernova gamma-ray limits of Refs.~\cite{Lella2023,Benabou2025} to a proton-coupled ALP with the same production rate.
(2) Repeat the survival calculation on full supernova profiles that include accretion and the ALP energy loss (Sec.~\ref{sec:formal_pns}), with the nucleon-velocity term in the hot rates.
(3) Replace the vacuum pions and $\Delta$s of Eq.~(\ref{eq:gpion}) by in-medium ones \cite{Fore2020}, and add bremsstrahlung on $\Delta$s in $\Delta$-admixed matter with couplings compatible with $2\,M_\odot$ stars \cite{Thapa2021,Parmar2025}.
(4) Extend the absorption calculation to dark photons and to ALPs coupled to muons \cite{Caputo2022} or electrons \cite{Lella2023}.

\begin{acknowledgments}
We thank Debajit Bose, Sougata Ganguly, Tanmoy Kumar and Nirmal Raj for correspondence on the $np$ channel and the induced photon coupling in Ref.~\cite{Bose2026ghosts}.
\end{acknowledgments}

\bibliography{refs}

\end{document}